%% file: STAR.tex
\documentclass{egpubl}
\usepackage{eurovis2026}

\usepackage{microtype}

\STAREurovis   % for EuroVis additional material

\CGFccby
\usepackage[T1]{fontenc}
\usepackage{dfadobe}

\biberVersion
\BibtexOrBiblatex
\usepackage[backend=biber, bibstyle=EG, citestyle=alphabetic, mincitenames=1, maxcitenames=1]{biblatex}
\makeatletter
\@ifpackageloaded{biblatex}{%
	\addbibresource{abbreviations.bib}
	\addbibresource{references.bib}
	\addbibresource{surveyed.bib}
}{\bibliography{abbreviations,references,surveyed}}
\makeatother

\electronicVersion
\PrintedOrElectronic
\ifpdf \usepackage[pdftex]{graphicx} \pdfcompresslevel=9
\else  \usepackage[dvips]{graphicx} \fi

\usepackage{egweblnk}
\usepackage{balance}
\usepackage{booktabs}
\usepackage{tabu}

\usepackage[table]{xcolor}

\definecolor{cvprblue}{rgb}{0.21,0.49,0.74}
\usepackage[role = cvprblue, skipempty, horizontal]{credits}

\usepackage{wrapfig}
\usepackage{parskip}
\usepackage{array}
\usepackage{float}
\usepackage{graphicx}
\usepackage{comment}
\usepackage{todonotes}

\usepackage{tabularray}
\usepackage{tabu}
\usepackage{booktabs}
\usepackage{adjustbox}
\usepackage{multirow}

\usepackage{etoolbox}

\usepackage[most]{tcolorbox}
\newtcolorbox{summaryboxLonni}[2][]{%
  enhanced,
  breakable,
  colframe=#2!60!black,
  colback=#2!8!white,
  boxrule=0.5pt,
  arc=2pt,
  left=6pt,right=6pt,top=10pt,bottom=6pt,
  before skip=6pt, after skip=6pt,
  fonttitle=\bfseries,
  title={Summary\ifstrempty{#1}{}{ \textbar\ #1}},
  overlay unbroken and first={
    \node[anchor=west, inner xsep=0pt, inner ysep=0pt]
      at ([xshift=8pt,yshift=-2pt]frame.north west) {\tcbtitle};
  },
  overlay middle and last={
    \node[anchor=west, inner xsep=0pt, inner ysep=0pt]
      at ([xshift=8pt,yshift=-2pt]frame.north west) {\tcbtitle};
  },
}

\providecommand{\tabularnewline}{\\}

\newcommand{\arxiv}{\raisebox{-0.2em}{\includegraphics[height=1em]{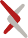}}}

\DeclareSourcemap{
  \maps[datatype=bibtex]{
    \map{
      \step[fieldsource=journal, match=\regexp{arXiv}, final]
      \step[fieldset=usera, fieldvalue=arxiv]
    }
    \map{
      \step[fieldsource=type, match=\regexp{arXiv}, final]
      \step[fieldset=usera, fieldvalue=arxiv]
    }
    \map{
      \step[fieldsource=publisher, match=\regexp{arXiv}, final]
      \step[fieldset=usera, fieldvalue=arxiv]
    }
    \map{
      \step[fieldsource=archivePrefix, match=\regexp{arXiv}, final]
      \step[fieldset=usera, fieldvalue=arxiv]
    }
  }
}

\AtEveryCitekey{\iffieldundef{usera}{}{\arxiv}}
\DeclareSourcemap{
  \maps[datatype=bibtex]{
    \map{
      \step[fieldsource=note, final]
      \step[fieldset=addendum, origfieldval, final]
      \step[fieldset=note, null]
    }
  }
}

\newcommand{\ie}{\emph{i.e.},}
\newcommand{\eg}{\emph{e.g.},}

\let\oldcitetitle\citetitle
\renewcommand{\citetitle}[1]{\oldcitetitle{#1} \cite{#1}}
\newcommand{\citetitleonly}[1]{\oldcitetitle{#1}}

\input{statistics}

\title{State of the Art of LLM-Enabled Interaction with Visualization}

\author[Brossier et al.]
{\parbox{\textwidth}{
        \centering
        M. Brossier$^{1}$\orcid{0000-0001-7653-9457},
        T. Isenberg$^{2}$\orcid{0000-0001-7953-8644},
        K. Schönborn$^{1}$\orcid{0000-0001-8888-6843}, 
        J. Unger$^{1}$\orcid{0000-0002-7765-1747},
        M. Romero$^{1}$\orcid{0000-0003-4616-189X},
        J. Björklund$^{3}$\orcid{0000-0003-0596-627X},
        A. Ynnerman$^{1}$\orcid{0000-0002-9466-9826}, and
        L. Besançon$^{1}$\orcid{0000-0002-7207-1276}
    }
        \\
    {\parbox{\textwidth}{\centering
            $^1$Linköping University, Sweden\hspace{10mm}%
            $^2$Université Paris-Saclay, CNRS, Inria, LISN, France\hspace{10mm}%
            $^3$Umeå University, Sweden%\\
        }
    }
}

\usepackage[notext]{stix} % to generate the correct dagger symbols

\begin{document}

\maketitle

\begin{abstract}

% Lonni + Mathis
We report on a systematic, PRISMA-guided survey of research at the intersection of LLMs and visualization, with a particular focus on visio-verbal interaction---where verbal and visual modalities converge to support data sense-making.
The emergence of Large Language Models (LLMs) has introduced new paradigms for interacting with data visualizations through natural language, leading to intuitive, multimodal, and accessible interfaces. We analyze \NumSurveyed{} papers across six dimensions: application domain, visualization task, visualization representation, interaction modality, LLM integration, and system evaluation. Our classification framework maps LLM roles across the visualization pipeline, from data querying and transformation to visualization generation, explanation, and navigation. We highlight emerging design patterns, identify gaps in accessibility and visualization reading, and discuss the limitations of current LLMs in spatial reasoning and contextual grounding. We further reflect on evaluations of combined LLM-visualization systems, highlighting how current research projects tackle this challenge and discuss current gaps in conducting meaningful evaluations of such systems. With our survey we aim to guide future research and system design in LLM-enhanced visualization, supporting broad audiences and intelligent, conversational interfaces.

\ccsdesc[500]{Human-centered computing~Visualization}
\ccsdesc[500]{Computing methodologies~Natural language processing}
\ccsdesc[300]{Human-centered computing~Visualization design and evaluation methods}

\printccsdesc

\end{abstract}

% \newpage
%%%%%%%%%%%%%%%%%%%%%%%%%%%%%
%% INTRODUCTION
%%%%%%%%%%%%%%%%%%%%%%%%%%%%%
\section{Introduction}

% \textcolor{red}{
% easy cosmetic TODOs:
%  [ ] add hyperref links in coding_summary
%  [ ] make sections the same color as in the corresponding column in coding_summary
%  [ ] check the references: remove duplicates, click DOIs, shorten journals
%  [x] justify refs in coding_summary
% }

%\textcolor{orange}{Lonni: 
% \begin{itemize}
%     \item This is a bit too much "history of vis" at the beginning, I suggest shortening this. 
%     \item You need to place the focus on the need to interact with visualization and to provide EDA both. Try to find a nice merge between those two in the intro.
%     \item There are papers for R, Tableau and Matlab I assume, would be nice to cite them or link to the websites at least. But a citation is always better.
% \end{itemize}
%}

Data visualization has long been a cornerstone of scientific inquiry and communication, enabling humans to externalize, explore, and reason about complex information. The earliest recorded maps, charts, temporal graphs, and other abstract representations of measurement date back to the 17\textsuperscript{th} century \cite{brief_history_datavis}. Beginning in the 1970s, the field leaped forward as computers became widely available. The authoring of a visualization was no longer limited by physical constraints or artistic skills.
Moreover, the ease of creating basic visualizations meant that authors became more efficient in their work. As a consequence, visualization developed beyond its original function as a means of communication to become an explorative tool for data scientists. The now-widespread field of Exploratory Data Analysis (EDA), which is based on gaining insight from data by means of statistics and visual representations, was concretized by Tukey in his eponymous book in 1977 \cite{tukey_exploratory_1977}. With the growing size and complexity of data sets, EDA is now an essential workflow for every data scientist, backed by both scientific literature and software engineering efforts (R, MATLAB, Tableau). 

Interaction capabilities are a central component of EDA systems and a major avenue of research in the visualization community \cite{besancon_spatial_interfaces_2021,perrin_vis_interact,yi_toward_2007}, specifically to ensure that visualization software helps users extract findings and understand their complex datasets \cite{besancon_spatial_interfaces_2021} and eventually adopt new visualization software, which remains an open challenge in the visualization community \cite{wang:hal-02053969}.
In the data science field, the combination of Machine Learning and Visual Interaction shortens the iteration loop of EDA through visualization recommendation \cite{survey_towards_vrs} and iterative refinement, lifting the burden of low-level and repetitive tasks from data analysts.
Interaction with visualization is, however, not limited to data science contexts.
In science communication contexts, interaction helps designers to foster engagement and tailor an explanation to a viewer  interest and level of understanding \cite{schonborn2016nano,Ynnerman:exp}.
Interactive visualization is therefore expanding beyond data science into educational contexts \cite{reaching_broad_audiences,reaching_broader_audiences}, including online learning materials, schools, and science museums.

The development of Natural Language Processing (NLP), particularly the advancement of LLMs, has opened new possibilities for interacting with data visualizations by leveraging \emph{visio-verbal} synergies. LLMs can interpret user queries expressed in natural language \cite{survey_nli_tabular_query_vis,survey_nli_for_dv}, generate relevant responses, and, in some cases, manipulate and update visualizations based on the interaction. Combining speech and visual cues is a natural way for typical humans to communicate. Systems that can leverage this combination have the potential to provide access to a slew of new synergetic interaction techniques, such as deictic gestures and speech (pointing and talking), contextualized explanations with visualization, and multimodal feedback. The ability of LLMs to generate text is not limited to natural language. With appropriate conditioning, LLMs are also proficient at generating code or structured data \cite{husein_large_2025}. LLMs can thus be used as intermediaries for natural language input.

While LLMs are certainly a ground-breaking technology, they also have several limitations that hinder their adoption in visualization systems.
Hallucinations are frequent when deviating from training data \cite{survey_ji_hallucinations_2023} and can affect trust in the system, convey false, incoherent, or misleading information \cite{detecting_misleading_vis,alexander_misleading_2024}. LLMs are also a source of biases due to their training data. Operating costs are high, which pushes for cloud-based approaches and induces dependence on private actors, environmental costs, and latency. More specifically to visualization, LLMs have a poor sense or spatiality, temporality, and relationships, making them essentially ``blind'' to the visualization \cite{mena_augmenting_2025,MenaSIGGRAPH}. This limitation can affect their understanding of queries, cause incorrect or inaccurate actions, and limit their ability to recover from such errors.

Considering both the growing potential of using LLMs to interact with visualization and their inherent limitations, we aim to survey the literature on the convergence of LLMs and visualization for interactive sense-making of data.
Specifically, by synthesizing this rapidly evolving landscape, our goal is to provide a comprehensive and structured view of how LLMs are influencing interaction with visualization, identify methodological and technical limitations, and propose research opportunities that can guide future LLM-enabled visualization system development and their evaluation.
One could question the relevance of a state-of-the-art report in an emerging and extremely rapidly moving field. We are, however, confident that the report will have important immediate impact by enabling researchers entering into the field to obtain a comprehensive overview of the area and thus significantly shorten the time to initiate novel research efforts in the domain. Our ambition is that the report will serve as a catalyst in the positioning of the visualization community in the rapidly evolving AI landscape. 

% \newpage
%%%%%%%%%%%%%%%%%%%%%%%%%%%%%
%% RELATED WORK
%%%%%%%%%%%%%%%%%%%%%%%%%%%%%
\section{Related surveys}

%\textcolor{orange}{Lonni: The goal of this section is to state which one is a survey clearly and how yours differs clearly from theirs. So after the first survey, explain what the difference is. }

Several research topics involving both natural language interaction and data science have been already been studied. Several surveys tackle a broader scope of visualization and automation, such as \textcite{survey_ml4vis} and \textcite{survey_ai4vis}, which both survey uses of machine learning-based automation across the visualization pipeline, but predate LLMs. Next, we focus on narrower topics in visualization. All topics presented are NLP techniques that existed before LLMs, but LLMs offer new opportunities, either in combination with, or as replacement of previous approaches. 

Data and information retrieval have been extensively studied and previously surveyed. In the topic of Natural Language Interfaces to Data (\textbf{NLID}) \cite{survey_nli_to_data}, database queries are automatically generated from natural language queries (\eg text-to-sql), facilitating queries such as ``retrieve the users which bought the article after January 7\textsuperscript{th}''. NLID involving LLMs was surveyed by \textcite{survey_hong_texttosql_2025}, but as \textcite{survey_liu_nli4db_2025} emphasize, LLMs techniques are not a one-fits-all solution, as they are more opaque and costly to operate. A related topic is Natural Language Knowledge Base Question Answering (\textbf{KBQA}), where users ask questions that a language model replies to supported by data extracted from a knowledge base. There is an overlap between data retrieval and knowledge retrieval, and both are employed in Retrieval-Augmented Generation (\emph{RAG}) by LLM systems, surveyed by \textcite{survey_zhao_rag_2025,survey_peng_rag_2025,survey_fan_rag_2025,survey_yu_rag_2025}.

Next, moving closer to our topic, Natural Language Visualization Recommendation Systems (\textbf{VisRec}) have also been extensively surveyed \cite{survey_towards_vrs,survey_auto_infographics}. VisRec is the task of automated visualization generation from natural language queries, such as ``plot the sales profit from 2010 to 2020 per product.'' It often involves NLID and sometimes KBQA \cite{survey_auto_infographics}, and approaches that rely on LLMs also exist \cite{wang_visualization_2025}. 

The previously mentioned surveys focus on essential tasks in visual data analysis, but our topic more specifically addresses Natural Language Interaction for visualization (\textbf{V-NLI}), in which natural language queries support users' interaction with an existing visualization \cite{survey_nli_for_dv,survey_nli_in_vis}. Here however, most previous surveys predate LLMs. Furthermore, these surveys focus on the perspective of a data scientist (not considering the perspective of broader audiences or accessibility) and the perspective of traditional visualizations or charts (not considering do\-main-spe\-ci\-fic visualization, which have unique problems and interest in LLMs). In other words, previous work focused on visualization \emph{authoring}, but not visualization \emph{reading} or interactive data analysis facilitated by LLMs. While there is a growing body of work focusing on using LLMs as an interaction assistant to facilitate sense-making and engagement with visualization, we could not locate an existing survey that has analyzed this particular gap in the literature. %Fewer works focus on how LLMs can be used as an interaction assistant to facilitate sense-making and engagement with visualization.

In this state of the art report we address this gap by conducting a systematic PRISMA-driven review of LLMs and visualization. Specifically, we study how LLMs contribute to and influence visualization workflows, classify LLM-sup\-por\-ted interaction modalities and visual representations, and examine domains where LLM-enabled interactive visualization systems are emerging. We also analyze LLM integration strategies, from prompting and fine-tuning to multimodal and agent-based architectures, and critically assess how current systems are evaluated, thus identifying potential methodological deficiencies. In particular, we concentrate on surveying the specific contribution of LLMs for \emph{interacting} with the user throughout the visualization authoring and reading process. We approach this challenge by decomposing the visualization process into tasks, following the taxonomy we describe in \autoref{sec:task-taxonomy}, and analyzing at each step the specific contribution of the LLM interactive component with the visualization. Our survey therefore distinguishes itself from past work by focusing on the user-LLM interface at different steps of the visualization and data-analysis pipeline.

% \newpage
%%%%%%%%%%%%%%%%%%%%%%%%%%%%%
%% METHODOLOGY
%%%%%%%%%%%%%%%%%%%%%%%%%%%%%
\section{Methodology}

\begin{figure}
    \centering
    \includegraphics[width=\linewidth]{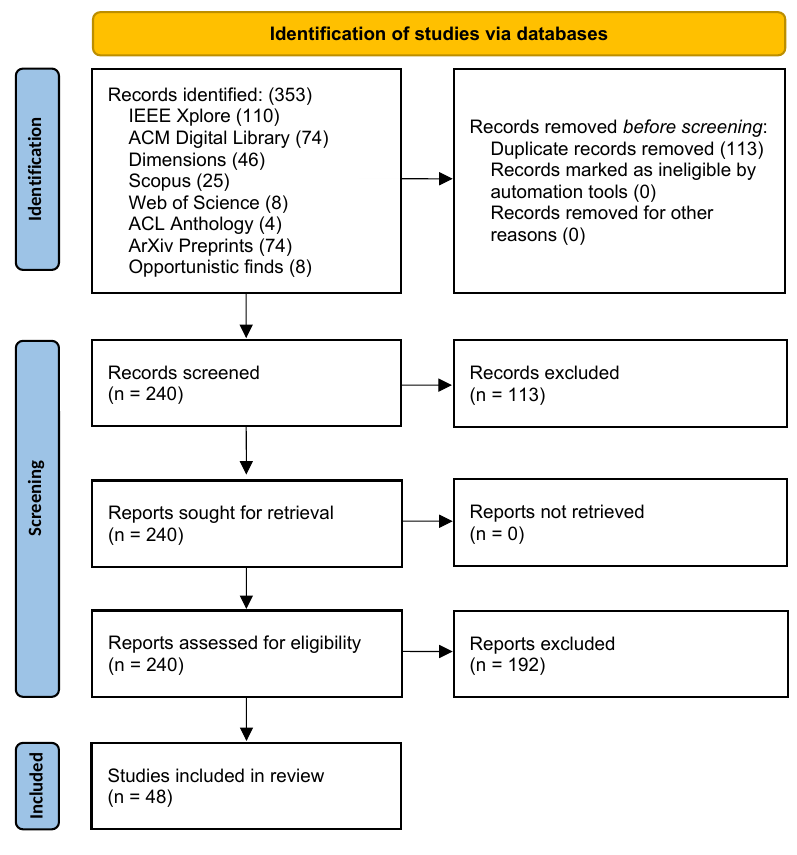}%
    \caption{PRISMA flow-chart of the paper selection process.}\vspace{-2ex}%
    \label{fig:prisma}
\end{figure}

In our methodology we followed the PRISMA-2020 guidelines \cite{prisma} as a structured approach to conducting systematic literature reviews with transparency and rigor. PRISMA includes a checklist to ensure transparency and completeness of the analysis, together with a flow-chart that reports the provenance of the surveyed studies (\autoref{fig:prisma}). We make all data that we produced for this survey available on OSF, doi: \href{https://osf.io/zhux9}{\texttt{10\discretionary{/}{}{/}nm8q}} (\href{https://creativecommons.org/licenses/by/4.0/}{CC-BY-4.0}). We also published an interactive browser for accessing the data on OSF.
% \begin{center}
%     \texttt{\href{https://llmvis.vlc.itn.liu.se}{https://llmvis.vlc.itn.liu.se}}
% \end{center}

\subsection{Search process}

We performed a systematic search in renowned databases and publishers connected to the visualization and natural language processing communities. We selected both multidisciplinary databases (Scopus, Dimensions, Web of Science); and more specialized databases (IEEE Xplore, ACM Digital Library, ACL Anthology), as well as arXiv preprints. We include all peer-reviewed academic publication types (books, journal publications, conference proceedings, workshops). We searched for papers from arXiv (non-peer-reviewed preprints) because LLM research is a particularly fast-mo\-ving field and many relevant submissions have not yet reached the status of peer-re\-viewed publication (see \autoref{sub:preprint_inclusion}). 
We executed the same following search query in all databases:

\texttt{visualization AND interact{*} AND ("large language model" OR LLM OR GPT OR transformer)},

where double quotes (\texttt{"}) indicate an exact text match and the star (\texttt{*}) is the wildcard symbol (matches any suffix). We ran the search in the title, abstract and keywords fields. This search returned \NumIdentifiedPapers{} papers, out of which \NumFilteredPapers{} were unique.\footnote{See the supplemental material for details on how to reproduce the search query.}

During the search, we found some discrepancy in the results returned by databases. For instance, using the arXiv API generated different results than the web interface. Likewise, on IEEE searching for \texttt{vi?ualization} returned different results matching ``visualization''. To address this issue, we simplified our initially more complex query to the one presented above, which yielded multiple false positives but increased our confidence that relevant papers were not overlooked. Our initial search query included clauses for targeting human-LLM interaction: \texttt{dialogue OR chatbot OR convers*}. Without these clauses, many of the returned papers turned out to be irrelevant, as they correspond to ML research papers with some visualization (\ie{} Vis4LLM, not LLM4Vis or LLM+Vis papers).

\subsection{Screening process}

The systematic screening in databases yielded \NumIdentifiedPapers{} records from the databases. All queries together returned \NumFilteredPapers{} unique and retrievable papers. In addition, we found \NumIdentifiedPapersOpportunistic{} further papers opportunistically, out of which we kept two. After this identification phase, we manually screened the papers to assess their relevance. We skimmed the papers to determine whether they met the eligibility criteria (\autoref{sec:eligibility}). We then assigned each paper to two contributors to tag it as ``relevant,'' ``irrelevant,'' or ``unsure.'' We assigned papers to a third contributor in case of disagreement or ``unsure.'' We kept papers with at least two ``relevant'' assessments, reducing the count to the \NumCodedPapers{} papers we include in this report. Later, we removed one retracted paper.\footnote{Our coding sheet and instructions are available in the supplemental material on our OSF repository.}

\subsection{Eligibility criteria}
\label{sec:eligibility}

Two contributors then skimmed each identified paper to determine its suitability. We established the following eligibility criteria:
\begin{itemize}
\item \emph{System}. The paper contribution must be a system or technique. We excluded surveys, case studies, and position papers.
\item \emph{LLM}. Must use a Large Language Model to accomplish or help accomplish a visualization task.
\item \emph{Visualization}. Must use data visualization. The visualization does not have to be the main contribution of the paper.
\item \emph{Interaction}. The user must have means of interacting with the LLM or the visualization to accomplish a task.
\end{itemize}

To avoid ambiguities, we clarified the following terms:
\begin{itemize}
\item \emph{LLM}: LLMs are natural language models with millions to billions of parameters, trained on terabytes of data and capable of general (non-specific) natural language generation. For the purposes of this survey, we consider that models qualify if they are based on the transformer architecture \cite{attention_is_all_you_need} or later improvements of this architecture. This definition includes multimodal models capable of receiving or producing media types other than text (commonly images, videos, or audio).
\item \emph{Visualization}: We chose the broadest definition of visualization, as any graphic representation of data. It encompasses traditional data visualization (tables, graphs and charts), as well as scientific and domain-specific visualization (maps, networks, 3D, etc.).
\item \emph{Interaction}: We follow the definition proposed by \textcite{perrin_vis_interact}, who define an interaction with visualization as ``the interplay between a person and a data interface involving a data-related intent, at least one action from the person and an interface reaction that is perceived as such,'' except that we replace the term ``person''
with ``actor'' such that LLMs are considered as an actor with ``intents'' that can initiate an action. We judged the question of ``perception'' though, from a human perspective.
\end{itemize}

\subsection{Preprint inclusion}
\label{sub:preprint_inclusion}

While it is commonly admitted that survey papers should focus on published and, as such, peer-reviewed contributions, we decided to include preprints from arXiv in this STAR. This decision is motivated by the extremely rapid pace of development in the field of LLM/AI research and the fact that many landmark papers in the field might not even get submitted for peer review. By excluding preprints we feared that we may not gain a complete picture of the current state of research. %That being said, we are also aware that peer review, even if far from perfect or a complete protection against fraudulent or low-quality research \cite{besanccon2020open,boetto2021frauds}, is usually helpful in providing some sort of quality filter. 
However, to ensure that we do not present all papers (peer reviewed or not) as providing the same strength of evidence \cite{besancon2021open}, we clearly separate them and indicate how many of those papers came from arXiv. We assessed these papers through the same process, but also eliminated those that we thought were not up to the quality standards that would be generally expected from our visualization publication venues. We discuss this issue more specifically in %\autoref{sub:quality_standards}
\autoref{sec:discussion}. We identified \NumIdentifiedPreprintsArxiv{} papers from arXiv, out of which we included \NumCodedPreprints{} in the analysis. We identify the preprints in our discussion and use references with the arXiv symbol (\arxiv) (\eg{} \cite{wang_visualization_2025}). With this method, we hope that we manage to clearly present the actual state of the art of the research in the field without sacrificing novelty or compromising rigor.

% \newpage
%%%%%%%%%%%%%%%%%%%%%%%%%%%%%
%% CLASSIFICATION
%%%%%%%%%%%%%%%%%%%%%%%%%%%%%
\section{Classification}

For our systematic survey we established a classification of LLM-enabled visualization systems. It is composed of four items: LLM-enabled visualization tasks, interaction modalities, visual representations, and application domain. Each item has several categories, and papers can fit in one or many categories of each item. \autoref{table:categories} summarizes the classification categories. We also analyzed the evaluation and implementation of each LLM-enabled visualization system.

\subsection{\textcolor{olive7!60!black}{LLM-enabled visualization tasks}}
\label{sec:task-taxonomy}

\begin{figure}
\centering
\includegraphics[width=0.5\linewidth]{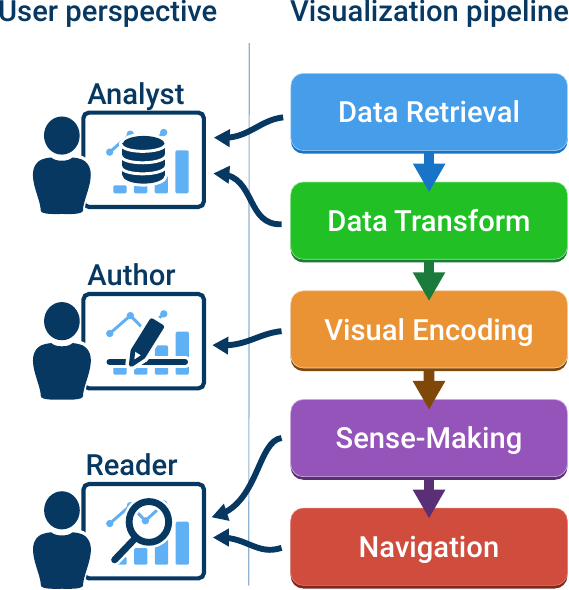}
\caption{Our typology of tasks in a visualization pipeline.}\vspace{-2ex}
\label{fig:tasks}
\end{figure}

In the first item, we categorize the visualization tasks performed by using the system. The purpose of this item is to allow us to identify what role LLMs can play in a visualization pipeline (\autoref{fig:tasks}). While numerous task taxonomies have been developed in the past \cite{chuah_semantics_1996,shneiderman_eyes_1996,ren_multilevel_2013,schulz_design_2013,rind_task_2016}, we decided to base our classification around the three core steps in the visualization pipeline, as described by \textcite{card_information_visualization_1999}: data transformation, visual mapping, and view transformation. We map these steps into three corresponding user roles: analyst, author, and reader. Shifting the perspective from tasks to user roles allows us to adopt a top-down vision with human input at the heart of the design. This perspective matches our goal of analyzing LLM roles in visualization. LLMs can automate, partially or fully, one or several tasks of either user role. This framing is inspired by existing taxonomies of visualization roles and tasks, such as the data state reference model \cite{tax_bertin_dsrm, tax_dsrm_vis}, which describes how information flows from raw data to visualization and perception, or by the knowledge generation model in visual analytics \cite{sacha_knowledge_generation_2014}, which describes visual analytics tight coupling between human input and the system.

To establish the tasks within the roles, we referred to narrower taxonomies focusing on one aspect of visualization: analytic tasks in visualization \cite{tax_amar_low-level}; tasks in graph visualization \cite{tax_lee_graph_vis}; tasks in collaborative visualization \cite{tax_grimstead_collab_vis}, and automated analytics tasks \cite{tax_domova_automation_va}. Closer to our topic, \textcite{tax_domova_automation_va} classify degrees of automation for different analytic tasks. Overall, we found that existing taxonomies related to automation or ML focus on analytics tasks, while our survey targets a broader scope that includes not only analytics but also visualization authoring and interaction. The closest to our classification is \citeauthor{survey_ml4vis}'s work \cite{survey_ml4vis}, who similarly break down seven visualization processes into three categories and include human components such as reading, profiling, interaction, and insight communication. Our classification adds the data retrieval category, which is not part of previous classifications, as LLMs are well suited for information retrieval and are a growing component of interactive visualization, which we found to be prevalent in the corpus. We also include a navigation task, which is particularly prevalent in 3D visualization, where changing points of view into the visualization is essential to grasp the full picture, and where fine view control is an interaction challenge \cite{keefe_reimagining_scivis_interact_2013,besancon_spatial_interfaces_2021,oviatt_multimodal_2002}.
We thus use the following final task categories:

\emph{Data retrieval}. Corresponds to fetching data from outside sources to produce a visualization. A common assumption in visualization is that the needed data is known and readily available. While this is true in most contexts, \ie{} if the data is produced for the analysis, or if the analysis is constructed around the data, in the case of explorative visualization, an analysis goal may emerge without specific data supporting it, in which case data retrieval is an initial step.

\emph{Data transformation}. Computations performed on the data prior to visual encoding, such as the seven low-level analytic tasks by \textcite{tax_amar_low-level}---retrieve value, filter, compute derived value, find extremum, sort, determine range, characterize distribution, find anomalies, cluster, and correlate.

\emph{Visual encoding}. This step maps data fields to visual representations. It includes choosing a visualization type and support, and mapping data to visual channels such as color, shape, and size.

\emph{Sense-making}. Tasks involving production and communication of insight backed by information.

\emph{Navigation}. When visualization is interactive, navigation regroups possibilities to adjust the view of the data during consumption of the visual media for exploration. Such actions include changing data ranges, filtering values, zooming and panning, and hover tooltips.

%% Figure: vis tasks pipeline
%% --------------------------
%% --------------------------

\subsection{\textcolor{cyan7!60!black}{Interaction modalities}}

The second item focuses on interaction modalities with the visualization system. LLMs provide a new communication channel with visualization, namely natural language, which is interfaced via textual and spoken interaction modalities. Moreover, with this classification we study the interplay between different interaction modalities, both LLM-enabled and otherwise. Indeed, humans communicate with a combination of elements that have different strengths including spoken language that describes abstract concepts, hand gestures that indicate spatial relationships, and facial expressions that convey emotions. Interactive visualization can be more effective by leveraging multimodality.

Like visualization, interaction remains an ill-defined topic that needs to be refined. We follow the definition proposed by \textcite{perrin_vis_interact}, where an interaction with visualization is defined as ``the interplay between a person and a data interface involving a data-related intent, at least one action from the person and an interface reaction that is perceived as such.'' This definition does not include LLM-based agents, which exhibit both machine characteristics and human characteristics, such as use of natural language. We adapt the definition and rename `person' with `agent,' such that LLMs become a third entity, in addition to the user and the visualization system, capable of using similar interfaces as the person.

There exists an extensive body of literature that examines interaction modalities from both VIS and HCI communities \cite{tax_bernsen_foundations_multimodal_1994,tax_augstein_interact_2019,yi_toward_2007}. Interaction modalities are often distinguished between ``input'' and ``output'' modalities \cite{tax_augstein_interact_2019}. Input modalities are initiated by a user, while output modalities provide sensory feedback \cite{benoit_audio-visual_2000}. Our classification focuses on input modalities, but we discuss LLM outputs in \autoref{sec:multimodal}. We included only input modalities used in the corpus. In \autoref{sec:opportunities} we discuss unexplored modalities.

Historically, use of textual and spoken interaction modalities in visualization was marginal \cite{keefe_reimagining_scivis_interact_2013}. These input modalities resurface with LLMs because they are direct input formats for LLMs. Some of the previous limitations of these modalities are now overcome by LLMs, but we do not yet know in what situations text and speech constitute superior modalities for interacting with visualization through LLMs. In other words, the most prevalent interaction input with LLMs is because it is more affordable, and not because it is more effective. We thus analyze uses of other input modalities enabled or supported by LLMs.
%
% \textcolor{orange}{Lonni: You're missing a few important vis references here to justify and relate your tasks to. }
%
% \textcolor{orange}{Lonni: Here I also expect to see you discuss the historical aspect of why voice/text was rarely used for interaction with data. In my STAR: "While voice input could also be considered, using voice for direct manipulation is generally discouraged [KI13] and it is seldom used alone. Consequently, voice input falls under our category
% of hybrid interaction paradigms." You can easily make the case that progress in NLP have made voice/text interaction for both data analysis and direct manipulation less so "absurd"}
% 
Finally, we settled on the following classification for the interaction modalities.

\emph{Textual natural language}, typed with a keyboard. The role of the interlocutor ranges from reacting to direct commands, answering question, to a symmetric, two-actor conversation.

\emph{Spoken natural language}, collected through a microphone.

\emph{Spatial interaction}. \textcite{besancon_spatial_interfaces_2021} identify four spatial interaction paradigms: tactile and pen-based, tangible and haptic, mid-air gestures, and hybrid. Our corpus does not contain any tangible interaction paradigm. % Hybrid interaction is mentioned in \autoref{sec:hybrid}.

\emph{User interface widgets}. The traditional keyboard and mouse paradigm on desktop include several standard interaction widgets, such as buttons, input fields, sliders, and pickers.

\subsection{\textcolor{yellow7!60!black}{Visual representations}}

The nature of data and its visual representation have high implications for the capability of an LLM to produce and interact with visualization. For instance, LLMs (including VLMs \cite{kamath_spatial_reasoning_2023}) can struggle with spatial representations \cite{wu_spatial_cognitive_2025} as they have poor spatial `reasoning' capabilities, compared to humans. Conversely, LLMs have a good `understanding' of clearly labeled tabular data as used in charts. Surveying visual representations (the graphical elements chosen to represent data) thus informs us on the strengths and weaknesses of LLMs in different visualization scenarios. Certain representations are more suited to a certain kind of data, \eg{} tabular data is often represented with 2D charts. Internally, there is a lot of possible variation in visual encoding, such as the nature of the chart. Design decisions matter for an LLM for both the generation of accurate and relevant visualizations as well as reading and interacting with existing visualizations.

Our typology of visual representations includes charts, spatial (maps, sca\-lar\discretionary{/}{}{/}vec\-tor fields), networks, ima\-ges\discretionary{/}{}{/}vi\-deos, and custom (in \autoref{fig:visreprs1} we show examples of visual data representations from the corpus):

\emph{Charts}. Traditional visualizations from tabular data such as scatterplots, bar charts, and pie charts. Typically produced by spreadsheet software and visualization software such as Vega-Lite \cite{vega-lite}.

\emph{Spatial}. Map, fields, or volumetric representations. While loosely related visualization techniques and intents, we group these together because they exhibit similar strengths and weaknesses for an LLM-based interaction.

\emph{Network}. Data formatted as graphs (having nodes connected by edges), representing networks (social networks, knowledge graphs).

\emph{Image}. While images stretch the usual definition of `data visualization,' they can easily be generated and manipulated by LLMs.

\emph{Custom}. Bespoke or rarely seen visualizations, not clearly fitting another category, fall into the `custom' category. Due to the lack of available training data for LLMs, they are harder to interact with.

\subsection{\textcolor{brown7!60!black}{Application domain}}
\label{sub:application_domain_justification}

% \textcolor{orange}{Lonni: Overall, you do a great job in this subsection I think and I would like all subsection to be as good as this one at least for the categories used.}

An application domain classification allows us to identify and compare do\-main-spe\-ci\-fic constraints that may affect the creation of LLM-en\-ab\-led visualization systems. Operative surgeons, \eg{} may seek speech control facilitated by LLMs due to the sterile environments on which they rely \cite{mewes_touchless_2017,besancon_spatial_interfaces_2021}. More generally, people working primarily with their hands may prefer voice input to interact with software \cite{aras}. In education, users may seek qualities of providing personalized explanation and engagement with a broad audience with poor visualization literacy \cite{reaching_broader_audiences}.

We coded papers using the Elsevier's three-tiered disciplines taxonomy \cite{tax_elsevier_disciplines} because it provided a sufficiently exhaustive list. Once we selected the papers, we narrowed it down to only five broad categories that we judged as sufficiently representative of the paper corpus, while being interesting for argumentation, as is common practice in STARs \cite{huang_va_embeddings_2023,star_sperrle_human_eval_2021}.

We thus classified as data science \& mathematics, medicine \& biology, physics \& engineering, and education \& social sciences. Among them, data science is a rather non-specific class: many papers fall into this category as they lack a specific domain focus. Some papers fall into two categories, yet due to our small data, we did not study synergies that may emerge from interdisciplinary examples.

\subsection{\textcolor{blue7!60!black}{System evaluation}}

Visualization research has a history of constantly rethinking evaluation methods to ensure that all important aspects surrounding visualization and its results are covered \cite{survey_lam_infovis_eval_2012}. Evaluation in visualization thus encompasses a wide breadth and depth of measures, from subjective assessment of systems (\eg{} NASA-TLX \cite{hart1988development} and SUS usability scales \cite{brooke2013sus}, aesthetics \cite{isenberg_beauvis_2022}, preferences \cite{besancon:hal-01436206}, readability \cite{Cabouat:2025:PPR}) to benchmarks for performances (\cite{tang-etal-2023-vistext,Wu_Benchmark}) and user performance (\eg{} task completion time, errors, perception tasks \cite{blascheck:hal-01851306}, intent \cite{Munzner} or tasks and strategies \cite{Srinivasan}). Consequently, we report on evaluation strategies in our survey. In doing so, we hope to establish a standard of best practices for future studies in the domain or highlight currently missing evaluation strategies that would allow us to better understand LLM-assisted interaction with data visualization.
We distinguish user evaluation and system evaluation. For the former, we report collected demographics (participant number, expertise level). For the latter, we report instruments, metrics, and benchmarks use.

\subsection{\textcolor{azure7!60!black}{LLM implementation}}

For researchers seeking to implement LLM components in visualization systems, we reviewed how the LLM integration was realized by the authors from a technological perspective. From the papers we collected and analyzed information about the integration of the language model into the visualization system.

We looked at which language models are used and what constraints guided the choice of a model. Then, we looked at the model instructions, in particular how system prompts are constructed, what prompt strategies are employed, what data goes in and out, and in what format. We also focused on the use of multimodal models capable of image and audio processing. Then we looked at mul\-ti-agent systems capable of delegating tasks to different specialized LLM agents. Finally, we discuss long-term memory, \ie{} means of retrieving of knowledge and data from external storage sources (e.g., retrieval-augmented generation and explicit memory modules) \cite{lewis2021retrievalaugmented,packer2024memgptllmsoperatingsystems,zhang2024surveymemorymechanismlarge} and learning over time from user interactions \cite{jin2025erarealworldhumaninteraction,zhang2025personalizationlargelanguagemodels}. These memory aspects are important to consider because they control whether the LLM is capable of providing information outside of the considered dataset and whether it has means to adapt and learn from users directly and thus provide increasingly more flexible interactions and better responses as the system is being used \cite{donyehiya2025naturallyoccurringfeedbackcommon,jin2025erarealworldhumaninteraction}.

%% Table: Coding items
%% --------------------------
\begin{table*}[!t]
\centering
\caption{Overview of the classification categories.}
\label{table:categories}
\begin{tabu}{X[0.2]X}
\toprule
coding items & categories \\
\midrule

visualization task &
\textbf{Data retrieval}: Extract data from databases \newline
\textbf{Data transformation}: Apply operations on data, such as sort, filter, aggregate, compare \newline
\textbf{Visual mapping}: Map data to a visual representation, including chart type, colors, layout \newline
\textbf{Sense-making}: Gain knowledge or insight from a visualization  \newline
\textbf{Navigation}: Select different information to be displayed \\
\midrule

visualization representation &
\textbf{2D Charts}: Traditional visualization charts from tabular data. Bar charts, scatterplots, pie charts \newline
% \textbf{XR}: Augmented Reality or Virtual Reality or Mixed Reality \newline
\textbf{Spatial data}: Map, fields or volumetric representations \newline
\textbf{Graph \& Network}: Graph representations \newline
\textbf{Multimedia}: Images and/or Videos \\
\midrule

user interaction modalities &
\textbf{Written natural language}: A text interface to have a conversation with an AI agent \newline
\textbf{Spoken natural language}: Voice interaction with an AI agent \newline
\textbf{Spatial interaction}: Hand-based interaction in mid-air or on touch-sensitive devices \newline
% \textbf{Sketching \& Drawing}: Free-form annotations \newline
% \textbf{Mouse selection}: Mouse highlights, box selection, lasso selection \newline
\textbf{GUI Widgets}: Mouse and keyboard interaction with buttons, drop-downs, sliders, \emph{etc}. \\
\midrule

application domain &
\textbf{Data science}, generally for systems that target no specific domain apart from visual analytics \newline
\textbf{Scientific subject domains} (Medicine, Biology, Physics, Chemistry, Engineering) \newline
\textbf{Education, Arts \& Social sciences} \\
\midrule

evaluation (data collection) &
\textbf{Participants}: Demography and number of study participants \newline
\textbf{Evaluation instruments}: questionnaire, interview, benchmarks, comparison studies, system logs  \newline
\textbf{Evaluation metrics}: user preference, task completion time, user recall, qualitative feedback, system accuracy, latency, and empirical observation \\
\midrule

LLM implementation &
\textbf{Model}: Model vendor and version \newline
\textbf{Training \& Fine-tuning}: Training process and data \newline
\textbf{Prompt engineering} : Structure of system prompts, use of few-shot, chain-of-shot \newline
\textbf{LLM data format}: Representation of data as input or output to the language model \newline
\textbf{Multimodal capabilities}: Training process and data \newline
\textbf{Multi-agent capabilities}: Architecture of systems involving multiple specialized agents\\
\bottomrule
\end{tabu}
\end{table*}
%% --------------------------

%% Table: Coding summary
%% --------------------------
\begin{table*}[!t]
\centering
\caption{All papers we surveyed. \textbf{\textcolor{azure7}{OpenAI}}: GPT model used. Shows only latest if multiple are used. \textbf{\textcolor{azure7}{num. participants}}: Number of participants recruited for user studies only. \textbf{\textcolor{blue7}{particip. expertise}}: Level of expertise of the majority of participants. N = novice, K = knowledgeable, E = experts, M = mixed.\label{table:coding_summary}}
\input{coding_summary} % contains the table generated by the spreadsheet.
% \raggedright
% \textbf{\textcolor{azure7}{OpenAI}}: GPT model used. Shows only latest if multiple are used. `?' if undisclosed. \newline
% \textbf{\textcolor{azure7}{num. participants}}: Number of participants recruited for user studies only. \newline
% \textbf{\textcolor{blue7}{particip. expertise}}: Level of expertise of the majority of participants. N = novice, K = knowledgeable, E = experts, M = mixed
\end{table*}
% --------------------------

% \newpage
%%%%%%%%%%%%%%%%%%%%%%%%%%%%%
%% SURVEY
%%%%%%%%%%%%%%%%%%%%%%%%%%%%%
\section{Survey of the state of the art}

We now present the survey of the state of the art in enabling interaction with visualization using LLMs. We first study the intersections between tasks and interactions from our classification and highlight gaps in the literature. Second, we focus on the visual representation and third, on the application domain in which those are used.

%% --------------------------
\subsection{LLM-enabled tasks in visualization systems}
\label{sec:tasks}

We first survey tasks in visualization systems, in conjunction with the interaction modality that enables it, as shown in \autoref{table:matrix_task_interact}.

%% Table: Vis Tasks categories
%% --------------------------
\begin{table*}[!t]
\centering
\caption{Survey of visualization tasks enabled by LLMs, sorted by interaction modality. The same study may appear in different cells.}
\label{table:matrix_task_interact}
\input{task_interaction_matrix} % contains the table generated by the spreadsheet.
\end{table*}
% --------------------------

\subsubsection{Task 1: Data and knowledge retrieval}

% \todo[inline]{include also (info retrieval): vist5,}
%\textcolor{orange}{Lonni: I'm not sure I get the difference between data and information. information is data, or? Perhaps explain this better but also in the previous section then.}
%\textcolor{orange}{Mathis: Renamed information to knowledge, hopefully it makes more sense. It's not the exact same meaning, but it works this way.}

Our first task is the retrieval of data and/or knowledge. Both concepts are related: they consist of fetching appropriate data or knowledge from outside sources to accomplish a specific task. Data and knowledge retrieval see a growing interest in interactive visualization systems. In typical scenarios, data are provided by analysts for specific analysis goals. With the increased availability of large, heterogeneous data corpora, however, the task of retrieving data for an analysis goal is becoming increasingly important \cite{mathiak_data_discovery_2023}, in particular for exploratory data analysis. Knowledge retrieval from databases is well-es\-ta\-blished, but LLM-based systems are now seen as a way to increase accuracy, reliability, and verifiability of LLM-pro\-duced information \cite{survey_zhao_rag_2025,survey_peng_rag_2025,survey_fan_rag_2025,survey_yu_rag_2025}.

In the literature, the retrieval task is automated by recommender systems, \ie{} systems capable of filtering and aggregating relevant data based on a user query \cite{survey_roy_recommender_2022}. Visualization systems capable of this task are correspondingly called visualization recommender systems (VRS) \cite{survey_agnostic_vrs,survey_vrs_strategies,survey_towards_vrs}. Within VRS, natural language approaches are popular due to their ease of use by humans \cite{survey_towards_vrs,survey_nli_tabular_query_vis,survey_liu_nli4db_2025,survey_nli_for_dv,survey_nli_to_data}. Such task automation predates LLMs, with, \eg{} rule-based natural language databases since the early 2000s \cite{survey_liu_nli4db_2025}. LLM-based approaches, however, have a better `understanding' of language nuances and context than the previously predominant rule-based techniques \cite{lu_tacit_2025,karanikolas_strengths_2025}, and are thus more likely to facilitate more explorative processes and provide access to non-expert audiences.  

We now focus on the means of interaction facilitated by the LLM for this retrieval task. Following our task-interaction matrix (\autoref{table:matrix_task_interact}), the next three sections focus on textual natural language retrieval, widget-based retrieval, and other approaches.

%\textcolor{orange}{Lonni: wait, here, we had more than three categories in the classification described in the section above. So why only three here, you need to be consistent.}
%\textcolor{orange}{I didn't fill in categories that didn't have relevant papers.}
 
\textbf{Retrieval with textual natural language.}
% "knownet meqa interchat"
% TODO chatweaver here?
%
In our corpus, several studies allow data queries in natural language, in particular for knowledge base question-answering (\emph{KBQA}), such as the work by \textcite{knownet,kosa}. As our survey focuses on systems providing interaction with visualization, we exclude systems performing only queries without further interaction. Instead, we direct the reader to the survey by \textcite{survey_liu_nli4db_2025} that addresseses this topic. We still include the work by \textcite{knownet,kosa} because they allow users to interact with the retrieved data.

A good example is \citetitle{knownet}, in which NL user queries are interpreted by a LLM and the LLM output is analyzed and linked to a knowledge graph (KG), storing semantic connections between terms. The authors identify three forms of knowledge: a procedural knowledge from human reasoning, knowledge learned by the LLM, and knowledge stored in the knowledge graph. They found that combining KGs with LLMs helps users reason about LLM-generated responses by navigating the structure of the KG. In particular, they show that their KG improved the assessment of subjective or uncertain phrasing by the LLM, such as \emph{``some research suggests that…''} or \emph{``[…] may have potential benefits….''}
\citetitle{meqa,chartgpt,chatgrid} employ a different approach. A Text2SQL agent \cite{survey_shi_text2sql_2025} translates NL questions into SQL queries. The authors then interrogate the database with the query and the response serves to augment the prompt of a second LLM agent. \textcite{chartgpt} explain that the approach is valuable in case of fine-tuning a LLM model specifically for the task is not an option, for example in presence of sensitive data, which the model is not allowed to learn from. \citetitle{chatgrid} does not perform any further data transformation after data retrieval, as SQL queries filtering and aggregation capabilities were found to be sufficient.

We did not find papers or approaches searching the web for information retrieval. This is particularly surprising since most consumer LLM chatbots (OpenAI ChatGPT, Microsoft Copilot, Google Gemma) have this capability built-in. Research in this direction exists outside of the visualization community. For example, in their survey \textcite{survey_deldjoo_genrecsys_2024} coined the term Gen-RecSys for retrieval systems with generative capabilities.
%\textcolor{orange}{Lonni: this is a good topic for more research in our future direction and discussion. Put a todo in the discussion/future research.}

To summarize, current research focuses on retrieving data using natural language from either knowledge graphs or SQL databases, but our survey highlights that, to the best of our knowledge, there have not been attempts made towards retrieving data from the internet, unstructured databases, or other forms of large data storage, such as books, specifically in visualization systems.

\textbf{Retrieval with widgets.}
% knownet meqa interchat
%
Widget interaction facilitated by LLMs support data retrieval in several papers. \citetitle{knownet} provides a timeline widget that allows revisiting previous queries, and provides clickable recommended queries. This widget provides users with an overview of the analysis progress and facilitates a quicker formulation of queries through recommendations. The authors propose that future work should implement more complex UI widgets. A tree view widget, for instance, would facilitate branching during the analysis. \citetitle{meqa} similarly provides clickable recommended follow-up questions, which trigger data retrieval.

\textbf{Unexplored interactions for retrieval.}
Natural language interaction has always been employed for data retrieval, predominantly in search engines. Yet, we did not find many papers which would directly address this task. We postulate that few authors focus on providing a complete visualization pipeline including data retrieval, due to the complexity of the task itself or the fact that it may fall outside of what would be considered interesting research contributions. Indeed, most current publications focus on a subset of the visualization pipeline (\autoref{table:coding_summary}). This focus may explain why contributions in this area appear sparse. Another possible explanation stems from the unpredictability of systems with vast retrieval capabilities. Granting access to more data to a LLM may increase the complexity and therefore, the unpredictability of the generated responses, which is already challenging to control with LLMs \cite{bench_bendeck_empirical_2025}. 
%  \textcolor{orange}{Lonni: unclear, what do you mean? Also, are you the only one making this point? Please link to more literature on the topic.}

%\textcolor{orange}{Lonni: Explain + back up with examples and literature.}
%\textcolor{orange}{@lonni, I don't have literature backing this claim, this is rather unsubstantiated, so I guess I'm making this up. I don't know how I should proceed here: drop the paragraph, or rephrase very carefully.}

Few systems provided retrieval mechanism from long-term memory. \textcite{autonomousgis} highlight in their limitations that ``every autonomous system needs a memory component to store the contextual and long-term information'' for iterative self-im\-prove\-ment. In addition, storing user-ve\-ri\-fied solutions to problems could increase the efficiency of the system for tasks not planned by system designers over time \cite{vizchat}. \textcite{c5} reverse the solution of long-term memory in visualization. They use visualization of conversation history to facilitate the recall of past dialogues for both the LLM and the human interacting with it, by retrieving and highlighting relevant past conversations during a LLM interaction. Information is stored and retrieved via topic classification.

\begin{summaryboxLonni}{cyan7}
Knowledge and data retrieval tasks see renewed interest with the development of LLM techniques, and some visualization systems started to adopt these developments. Current research, however, mainly focuses on natural language retrieval from knowledge or information databases. Retrieval is performed most often via SQL queries (Text2SQL) or via embeddings and similarity search with language models. Interaction modalities that trigger retrieval are mainly natural language text and GUI widgets. None of the surveyed studies employed other means of interaction to trigger retrieval. Digital assistants, \eg{} frequently allow asking questions out loud that trigger a search in a knowledge database. Furthermore, none of the studies support searching from external data sources, such as performing web searches or crawling the file system. Yet, consumer LLM products easily support these features but its inclusion may further increase the unpredictability of LLM-generated responses.
\end{summaryboxLonni}

%% Figure: visualization representations
%% --------------------------
\begin{figure*}[t]
\includegraphics[width=\textwidth]{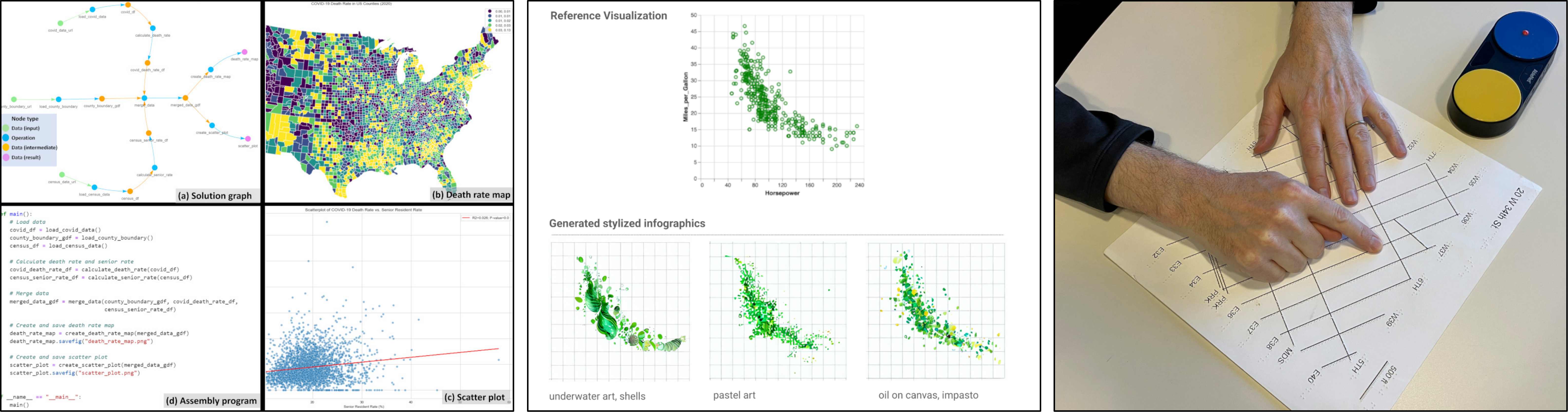}\vspace{-2ex}
\caption{
Representative visualization kinds in the survey. Left to right:
\textbf{(1)} Solution graph and map visualization, in \citetitleonly{autonomousgis} by \textcite{autonomousgis}. 
\textbf{(2)} Style transfer applied to generated infographics, in \citetitleonly{lida} by \textcite{lida}. 
\textbf{(3)} NLI assistant for assisting chart reading for BLV people, in \citetitleonly{mapio} by \textcite{mapio}. 
Images courtesy of the respective authors.
}%
\label{fig:visreprs1}\vspace{-1ex}
\end{figure*}
%% --------------------------

\subsubsection{Task 2: Data transformation}

Once the data are queried, they are transformed through mapping, filtering, and aggregation. These transformations process raw data from observations into a visualization abstraction \cite{tax_bertin_dsrm}. Basic transformations can be handled in the query phase by a centralized system (\eg{} SQL databases), but more complex data analysis tasks or interactively run transforms are implemented in the application via dedicated interactions or with user-writ\-ten code for bespoke tasks. We focus on interactions that trigger data transformations.

\textbf{Data transformation with textual natural language.}
% "autonomousgis chartgpt dataformulator nl4dvllm lida vist5 waitgpt meqa interchat"
%
Internally, there are different approaches taken by LLM-enabled systems to implement data transformation from natural language prompts. The first approach, adopted by a majority of papers, consists of generating code that has access to the dataset and produces a visualization using a specific framework. For instance \citetitle{lida,waitgpt,meqa,interchat} generate Python code to produce a chart matching the query, with little restrictions on the code produced. These systems do not distinguish between the data transformation and the visualization phase, so the code produced is responsible for both. \textcite{lida} augments the prompt with a structural and semantic description of the dataset. This grounding information reduces the likelihood of hallucinations. Furthermore, to help produce correct code the authors, \citeauthor{lida} split the generation into three phases: \emph{scaffolding}, where the program skeleton is generated; \emph{generation}, where the code logic is produced, including code outputting the visualization; and \emph{execution}, which executes multiple generated scripts and arbitrates on the best ones. \citetitle{waitgpt} produces unrestricted Python code, but is accompanied with an interactive graph visualization of the data transformations to the side, which gives the user insight and control over the generated code. In \citetitle{meqa}, most of the data transformation is done in the retrieval phase via SQL query operations. Then, unrestricted Python (Plotly) code is produced. \citetitle{interchat} produces unrestricted JavaScript (D3) code.

A different approach was adopted for the systems \citetitle{nl4dvllm} and \citetitle{vist5} by producing a structured, declarative data format that is interpreted by a visualization software. \textcite{vist5}, \eg{} produce a Vega-Lite specification of charts, which includes filtering and aggregation operations.
% \citeauthor{nl4dvllm} generate a structured JSON object \todo[inline]{I'm not sure if nl4dvllm actually does transform. Is it just encoding? \@ Tobias can you check???}, while 

Finally, a last approach to Task~2 is iterative. In \citetitle{autonomousgis,chartgpt,dataformulator}, the LLM is operated with a chain-of-thought prompt that produces a solution plan for a user query, involving a composition of elementary operations. \citetitle{chartgpt} applies elemental operations in fixed order (select columns, filter rows, aggregate). \citetitle{dataformulator} derives concepts (transformations with semantics) and combines them to reach the goal. In contrast, \citetitle{autonomousgis} produces Python code for the individual operations, then combines them into a final script to complete the task (\autoref{fig:visreprs1} \textbf{(1)}). \textcite{autonomousgis} highlight that single-step code generation would have worked yet the indirect approach (chain-of-thought, divide-and-conquer strategies) yields better results for complex tasks and produces verifiable outputs due to the solution graph visualization.

\textbf{Data transformation with spoken natural language.}
% "interchat genius"
%
One study, \textcite{interchat}, leverages spoken utterances for data \mbox{transformation}. It allows users to input transformation queries as either text or speech through the microphone. However, the authors do not compare the two modalities in their studies, nor comment on actual usage. They note that ``more complex modalities—such as advanced touch gestures, voice commands, and movement-based interactions—remain underexplored.'' This sentiment is surprising, as speech is used for other tasks (mainly navigation and explanation). Some authors argue that spoken interaction is best suited for a broad audience target \cite{voice}, or in operational settings \cite{word2wave}, such as work tasks requiring the use of hands \cite{aras}. But data transformation is a task primarily for visualization authoring and analysis, which is typically done sitting at a desk, where a speech modality may be a nuisance or a concern in shared office spaces \cite{jahncke2012open,Zulfikar}, or because the modality does not integrate well with the current workflow of different experts \cite{besancon:tel-01684210,isenberg:hal-01095454,wang:hal-02053969,wang:hal-02442690}. Nevertheless, further work is needed for some data analysis applications. For instance, all XR applications we surveyed employ a speech modality for navigation and/or explanation, but do not perform data authoring or analysis \cite{handtracking,genius,humanatlas,handmol,3dllava}.

\begin{figure*}[t]
\includegraphics[width=\textwidth]{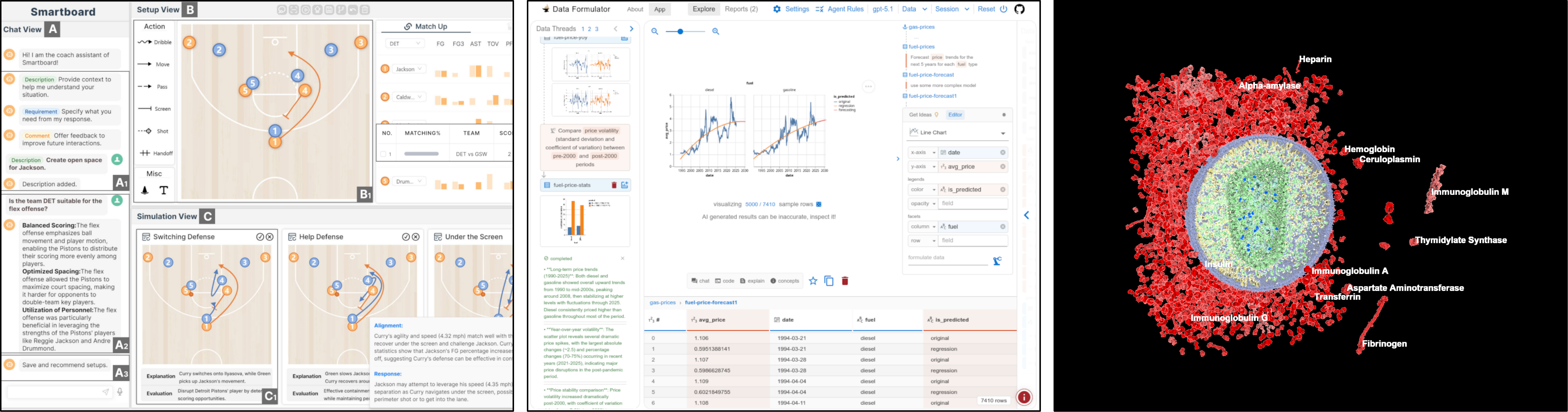}\vspace{-1ex}
\caption{
Representative visualization kinds in the survey.  Left to right:
\textbf{(1)} Generated player trajectories in basketball, in \citetitleonly{smartboard} by \textcite{smartboard}. 
\textbf{(2)} Building visualization from modular and interactive data concepts, in \citetitleonly{dataformulator} by \textcite{dataformulator}. 
\textbf{(3)} 3D molecular visualization and navigation, in \citetitleonly{voice} by \textcite{voice}. 
}%
\label{fig:visreprs2}\vspace{-1ex}
\end{figure*}

\textbf{Data transformation with widgets.}
% "chartgpt dataformulator vist5 waitgpt meqa interchat"
% remove lida
%
%Based on our survey (\autoref{table:matrix_task_interact}) 
Nine out of \NumSurveyed{} studies employ GUI widget interaction for data transformation. Basic interaction with mouse-ac\-ti\-va\-ted UI widgets is often implemented in visualization frameworks used for quick filtering and navigation of data using built-in widgets (\eg{} D3.js, Vega-Lite, or Plotly). These interactions, however, are not facilitated by LLMs. Other systems leverage traditional GUI widgets to craft and customize LLM prompts. Indeed, while natural language is efficient for conveying complex and abstract intent, it can be more verbose and error-prone than traditional GUI inputs for simple mechanical tasks \cite{chartgpt}. \textcite{chartgpt}, \eg{} break down the visualization recommendation into subtasks executed by the LLM. Data transformation steps include selecting columns, filtering rows, and computing aggregates. Due to abstract features of language, the authors explain that one query leads to a tree of possible interpretations. The LLM produces multiple charts and data transformation steps corresponding to different interpretations. The user can then both refine the natural language query, or modify directly the transformation operations using GUI widgets. \citetitle{dataformulator} operates at a lower level, where data transformations are encoded by ``data concepts'' derived from other concepts with a transformation intent. The visualization author provides input concepts and optional exemplar values in GUI, and the transformation intent in natural language. The LLM then generates the body of a function executing the intent with the source concepts as input (\autoref{fig:visreprs2} \textbf{(2)}).

\textbf{Unexplored interactions for data transformation.}
We did not find clear examples of data transformation driven by another interaction than natural language or widgets. While \textcite{genius} mention the use of hand gestures in mixed reality, this workshop paper does not provide much detail that would allow us to infer how these are actually used and for what purposes. We decided, nonetheless, to mention it here as a paper that potentially uses hand gestures for data transformation, since the manuscript seems to hint at this possibility.
Future work could focus on implementing data manipulation operators with LLMs via different input modalities such as mid-air gestures, sketching, or tangible interaction. Such setups have been studied without LLMs \cite{piper_illuminating_2002,}. LLMs could facilitate such interaction by providing the reasoning that translates the interaction raw input into data transformation primitives.

\begin{summaryboxLonni}{cyan7}
Most of the work we surveyed focuses on textual natural language, where LLMs either generate unrestricted code, produce structured declarative specifications, or follow iterative, multi-step reasoning to compose transformations. Fewer systems explore spoken language, where we see potential for data transformation beyond its anecdotal use in navigation and explanation tasks. GUI widgets are the most common modality, often complementing LLMs by constraining or refining transformations, combining the expressiveness of natural language with the precision and reliability of direct manipulation.
\end{summaryboxLonni}

%% Figure: visualization representations
%% --------------------------
%% --------------------------

\subsubsection{Task 3: Visual encoding}

Once data are extracted and transformed, the following visualization pipeline step is to choose a visual data mapping. This step involves choosing a visual representation (\eg{} the chart type) and the properties of visual components (colors, layout, textual information). An automatic generation of a visualization from data is called visualization recommendation. LLM-based VRS produce either imperative code or declarative descriptions of the output visualization. State-of-the-art LLM-based systems perform well at simple visualization recommendation tasks for most common 2D charts types, as there is ample training data available. \textcite{khan_evaluating_2025} report a success rate of 95\% for GPT-4o (best performer) for simple chart generation tasks in Python. They perform worse with precise customization requests, such as specifying colors, sizes, labels, \emph{etc.} The performance with Vega-Lite code generation was significantly lower, hinting that widespread availability of code examples is a key factor in LLM's visualization code generation performance. 

\textbf{Visual encoding with textual natural language.}
% "autonomousgis ava chartgpt nl4dvllm lasek lida vist5 waitgpt meqa interchat"
%
Similarly to data transformation, several systems employ LLMs for either generation of code \cite{lida,waitgpt,interchat} or declarative specifications \cite{nl4dvllm,vist5}. These systems produce only regular 2D charts. Depending on the desired level of automation, a recommender system automates the choice of a chart type, represented data features, and visual encoding parameters for a given query. Some systems provide interactive refinement of the visual mapping after an initial generation, with follow-up queries or other interaction inputs (which involves a dialog history for the LLM).

Although common VRS generate charts via analytic specifications (imperative or declarative code), we found one paper, \citetitle{lida}, which applies a post-processing pass to apply style transfer to charts using a text-conditioned image-to-image model (see \autoref{fig:visreprs1} \textbf{(2)}), for domain alignment and communicative purposes. This approach, however, raises questions about the balance between aesthetics and the accuracy of presented information \cite{wood2022beyond}.

Beyond 2D charts, visual encoding is also needed in 3D visualization. \citetitle{3dllava} works on point clouds of 3D scenes and uses a 3D-LLM \cite{hong_3d-llm_2023}, \ie{} a multimodal model capable of handling point clouds as input. The system is capable of segmenting regions of space based on natural language prompts, such as ``highlight the bathroom sink.''

\textbf{Visual encoding with spoken natural language.}
% "interchat genius word2wave"
%
Using voice commands to describe the visual layout facilitates quicker iterations during the design and prototyping phase \cite{kim_vocal_2019}. \citetitle{word2wave} works on the basis of the data points dictated out loud, which are displayed on the screen. For quicker feedback, the authors employ a fast, custom word-to-word model trained on a dataset generated with human-LLM collaboration.

\textbf{Visual encoding with widgets.}
% "ava chartgpt lida vist5 waitgpt meqa interchat"
%
Once an initial visualization is created with a natural language prompt, tweaking the visual encoding can be quicker and easier to do with GUI widgets---using a color picker, \eg{} can be more precise than trying to ask an LLM for a specific color. In \citetitle{interchat}, a system that offers both natural language and various GUI interactions, users tend to use natural language in the exploration phase, when they were uncertain of the exact requirements, but used direct interaction otherwise. Once the initial chart is generated, \citetitleonly{interchat} provides a means to iterate on the design or explore data by altering selected data inputs directly within the conversation history. The authors explain that users benefit from progressively specifying intent through a combination of language and visual selection, while direct language input is often preferred for better-defined tasks.
In \citetitle{waitgpt}, the LLM produces visualization by generating a flow diagram, which combines visual mapping and encoding primitives to produce a final chart. The flow chart doubles as an interactive GUI widget, in which users can easily tweak the visual encoding by editing input forms.

\textbf{Unexplored interactions for visual encoding.}
% genius
Apart from natural language and widgets, we did not find other interactions to handle visual encoding. This fact may be explained by a lack of relevance in studying other interaction modalities to handle visual encoding.

\begin{summaryboxLonni}{cyan7}
Our survey showcases several visualization recommender systems that accept natural language input to ultimately translate high-level user intent into concrete visualization encodings, including chart type, visual mappings, and layout decisions. In most of our surveyed systems, this process is inherently iterative, and users refine the generated visualization through follow-up queries.
Some papers mention the benefits of combining NL input with direct manipulation (\eg{} \cite{interchat,waitgpt}), especially for smaller refinements to tweak the visual output. However, these remain preliminary and interesting findings that would tend to suggest that direct natural language queries would be most effective when integrated with, rather than replacing altogether, traditional interaction techniques.
%One paper in particular \cite{interchat}, describes that multimodal interaction supports different steps of analytical needs. The authors explain that users benefit from progressively specifying intent through a combination of language and visual selection, whereas for better-defined tasks, direct language input is often preferred. 

A small set of the surveyed systems explores vision-based validation and refinement of the generated visualizations. Approaches such as \citetitle{3dllava} and \citetitle{visualizationary} incorporate computer vision models or heuristics to analyze visualization outputs and guide users through iterative improvement, yet more work seems needed on methods to ensure correctness.
Similarly, despite the growing multimodal capabilities of recent models, the use of style transfer for visualization is rare. With limited exceptions (\eg{} \cite{lida}), it appears that current research does not leverage generative models for aesthetic or communicative adaptation, highlighting an opportunity for future research, provided that data fidelity and visual integrity are preserved (\eg{} \cite{kouts_lsdvis_2023}), or, perhaps, not preserved for specific intents and purposes (\eg{} \cite{wood2022beyond}).
\end{summaryboxLonni}

\begin{comment}
Mathis' notes from which Lonni wrote the summary.

all visualization recommenders (input NL query -> ... -> output visualization) must do visualization encoding, to choose the chart type, colors, and layout in case of infovis. User can iteratively ask for style change / refinements in most papers.

when direct interaction with widgets is more efficient, offering choice between NL input and GUI input proved effective by \cite{interchat}: "When users are uncertain of exact requirements and need progressive exploration, multimodal interactions offer flexible and intuitive means to delve into the dataset. For example, to compare open and close prices on days with high trading volume, P5 generated a bar chart for trading volume, selected the bars with high volume, and instructed the system to generate a line chart. Conversely, for well-defined tasks like single-intent queries, participants often preferred direct language input, perceiving multimodal interactions as unnecessary."

vision models can be used to validate and refine (iterative loop) the output vis, as shown in \cite{3dllava} and \cite{visualizationary} (in this case, not a vision model, just computer vision algorithms and heuristics). But underexplored.

using style transfer capabilities of multimodal models is not explored, apart from \cite{lida} but could be interesting for aesthetics and comm purposes, while we need to be careful to keep the data intact. (cite lsdvis?)
\end{comment}

\subsubsection{Task 4: Navigation}

This task corresponds to view manipulations to obtain different viewpoints in the data. It is particularly prevalent in 3D scenes \cite{besancon_spatial_interfaces_2021}, in particular with mid-air hand gestures \cite{kostic_exploring_2024}, but operations such as zooming, dragging, and rotating (\eg{} on maps) are also relevant in 2D. On the one hand, GUI-based navigation (buttons, sliders, selection; \ie{} direct interaction) is well suited for fine-grained transformations that are tedious to describe with concepts (\eg{} zooming, moving a camera in 3D space, selecting an unlabeled data entry) \cite{masson_directgpt_2024}. On the other hand, NLI is best used for easily describable transformations. Ultimately, combinations of GUIs with natural language input facilitate an efficient navigation that combines an action description with direct manipulation \cite{su_natural_2021,shen_prompting_2025} (\eg{} pointing and asking ``show me this'').

\textbf{Navigation with textual natural language.}
% "datadive graphologue insightlens vist5 vizability vizchat voice interchat smartmlvs chatgrid"
% \todo[inline]{\citetitle{chatgrid,smartmlvs} has NL interaction but not for nav}
%
\citetitle{interchat} provides natural language view transformation combined with UI data selections. For instance, a user might perform a box selection with the mouse and utter ``zoom on this selection.'' Though, all interactions are available as either natural language queries (text or speech) or direct manipulation with the mouse.

\citetitle{voice} facilitates navigation in 3D space, around molecular models  (see \autoref{fig:visreprs2} \textbf{(3)}). The authors split navigation operations into two categories: camera adjustments and visual effects. Due to the hierarchical structure of the data being visualized and the complexity of specifying certain navigation tasks (such as camera movement), the authors define different navigation environments that limit available operations based on the context. Overall, the set of available transforms is carefully controlled at all times to prevent the LLM from steering the scene into an unwanted state. \citetitleonly{voice} also relies on a multi-agent LLM architecture, which has two specialized agents for navigation: the \emph{explorer} to take care of local transformations and animations and the \emph{pilot} to handle scene transitions. Finally, the system facilitates indirect navigation, \eg{} to explain a concept to the user, a manager LLM may build a sequence of view points to visit, without the user explicitly asking for it.

\textbf{Navigation with spoken natural language.}
% "voice interchat"
% add vist5, aras
%
Four papers \cite{voice,interchat,vist5,aras} rely on textual or spoken natural language input for navigation. A lot of literature covers navigation interaction in visualization \cite{bolt_put-that-there_1980,kim_datahand_2021,srinivasan_natural_2017,aurisano_show_2017}, and speech interaction for navigation provides benefits for accessibility to people with disabilities \cite{zhu_speech-based_2009,aziz_voice_2022}, to reduce fatigue in AR/VR use \cite{su_natural_2021}, and in collaborative work \cite{leon_talk_2025}. The papers we surveyed do not provide strong justifications for using speech input, nor do they seem to evaluate it when compared to textual input. \textcite{voice} emphasize that spoken natural \mbox{language} can increase user engagement and satisfaction, reduce \mbox{latency}, and increase intuitiveness. \textcite{vist5} explain that ``typing-based chat is very impractical, as it is annoying to switch back and forth between the keyboard and the screen'' and \textcite{aras} state that NLI is ``an optimal interaction method and reduce surgeons’ cognitive workload.'' Given the discrepancy between the little use of LLMs and the large evidence of the benefits of speech-based input for navigation, we suggest that there is an important research gap to fill.

\textbf{Unexplored interactions for navigation.}
We found little interest in navigation with widgets or spatial interaction. Papers tackling 3D spatial data often rely on mid-air gestures (with or without controllers) for navigation \cite{genius,avatar}. Our corpus contains only few AR/VR applications, and they do not use gestures combined with LLMs. Nonetheless, a clear research avenue exists in combining hand gestures with other modalities powered by LLMs, particularly with the recent advent of low-latency vision-capable models \cite{zhang_eyes_2025,chen_videollm-online_2024}, which may, in the near future, be capable of responding to arbitrary gestures in the same way that speech LLMs now `understand' spoken language. We found one paper, \citetitle{interchat}, that combines speech and mouse input to multimodal interaction. When performing a box or lasso selection on a chart, \citetitleonly{interchat} provides a selection context to the LLM prompt to facilitate deictic interaction such as ``Zoom into this and move to that place,'' where ``this'' and ``that'' correspond to previous mouse selections.

% gestures (handtracking)

\begin{summaryboxLonni}{cyan7}
Voice-based navigation is particularly valuable in contexts where keyboard and mouse input is impractical or unavailable, such as virtual and augmented reality or operational, hands-busy environments, where users often rely on direct spoken commands to control navigation and views. In contrast, in conventional 2D desktop environments, navigation is generally well supported by direct manipulation with mouse and keyboard, which likely explains the more limited adoption of speech-based navigation in such settings. Multimodal navigation that combines speech with complementary input has been studied extensively in pre-LLM systems \cite{besancon_spatial_interfaces_2021}, but only a small subset of papers used voice combined with other input modalities. LLM-augmented visualization could thus lead to more interesting applications. The closest examples combining LLM-based input and other navigation modalities include systems such as \citetitle{interchat} and \citetitle{inksight}, which combine natural language with GUI selections or free-form lasso input. Extending these approaches to additional input modalities, such as hand gestures or pointing in VR, represents a promising direction for future work, particularly as LLMs become more capable of integrating heterogeneous interaction signals. Yet, more research needs to be done to solve LLM-blindness for the specific case of data visualization (\eg{} \cite{mena_augmenting_2025,MenaSIGGRAPH}).
\end{summaryboxLonni}

\begin{comment}
Mathis' points before Lonni's version of the summary 

enabling navigation with voice is valuable for applications where keyboard/mouse are not approriate, such as vr/ar and operational work. Typically direct commands such as show this/that are used in this case.

in 2d there are fewer uses of navigation navigating in 2d is easier -> direct navigation is better/sufficient \todo{this is off my ass, let's source it}

synergetic navigation e.g. combining hand gestures, speech commands have been studied in the past, but we haven't seen a wide body of work doing this with LLMs. The closest is \cite{interchat} and \cite{inksight} where GUI selections + free draw (lasso) are fed to the LLM and allow combined interactions (example from interchat: "show me the load levels in this location", where "this location" is hand drawn). there is opportunity for future work with different input modalities such as gestures. In VR, what is done by \cite{interchat} and \cite{inksight} could be reproduced with hand controllers ("what am I pointing at?")
\end{comment}

\subsubsection{Task 5: Sense-making}

Sense-making tasks correspond to gaining knowledge or insight from visualization. Although LLMs are potentially beneficial tools for providing explanations tailored to context and users \cite{tailormind}, they must be used carefully. \textcite{koonchanok_trust_your_gut_2025,wang_aligned_takeaways_2025} found that insight produced by LLMs is not identical to insight that would be produced by human analysts. In many cases, a LLM performs worse than an expert in generating insights from a dataset. LLM-generated explanations are also associated with risks of hallucinations \cite{survey_ji_hallucinations_2023}, LLM sycophancy (\ie{} agreeing with user's preconceptions, being ``people pleasers'') \cite{fanous_syceval_2025}, and even deception \cite{greenblatt_alignment_2024,park_ai_2024}. 
Of note, some of the papers we found make use of LLMs to, \eg{} assist with structuring insights. For instance, \textcite{Pooryousef_coll_gen_ai} use generative AI to allow forensic experts and pathologists to automatically structure autopsy reports for court cases as they use immersive autopsy and visualization software \cite{Pooryousef_forensics}. We decided against including those in our analysis since they do not directly use LLM to interact with the visualization.

\textbf{Sense-making with textual natural language.}
% "aicommentator datadive graphologue insightlens knownet cellsync vizability vizchat voice"
%
% \todo[inline]{here i'm tempted to repeat some of the stuff in data retrieval section, since they serve the same purpose.}
%
Sense-making through textual natural language is primarily enabled through dialogic question-answers with a LLM assistant. Out of the box, LLMs can provide explanations to general questions through their internal knowledge. This is insufficient when domain-specific knowledge is required, in which case fine-tuning and retrieval augmented generation are efficient approaches.  

\citetitle{aicommentator} provides commentary with embedded visualization for viewers of football games. The system embeds visualization into the video feed, such as tracking the ball and player, and showing player statistics. In interactive mode, users can query these visual augmentations through natural language texts such as ``Track [player].'' In commentary mode, the system provides these visual highlights automatically during visioning. Study participants showed a slight preference for the interactive mode. Less knowledgeable participants, however, did not ``know what to ask'' in interactive mode, while knowledgeable participants enjoyed asking questions.

\textbf{Sense-making with spoken natural language.}
% "aras voice ephemera avatar"
%
Two papers provide explanations for spoken queries through speech synthesis, \citetitle{voice} and the work by \textcite{avatar}. Both systems employ a three-step setup, with first a speech-to-text model, fed to a textual LLM, whose output is then synthesized to speech with a text-to-speech model. Since 2024, end-to-end speech LLM models are available to the public, which have reduced latency, accuracy and can respond to intonations patterns. We thus strongly suspect that more speech-based dialogic systems will see the light of day in the coming years. To provide contextualized explanations, \citetitle{voice} combines a spoken explanation with camera motion paths. However, the LLM prompt has limited understanding of the visual layout (\eg{} elements in focus and color mappings), which prevents users from asking visual-dependent questions (``what is the red spot?''). Future work, such as that of \textcite{MenaSIGGRAPH}, can focus on enriching LLM prompts with contextual visual metadata \cite{mena_augmenting_2025,MenaSIGGRAPH}.

\citetitle{articulatepro} however, listens to users but does not reply with speech synthesis. Instead, it generates appropriate visualization to user queries. The system is continuously listening to multi-user conversation. It generates visualization when a user formulates either an explicit request or proactively. For the proactive case, the system contains a module to detect situations that require the generation of an explanatory visualization. For instance, a user may say ``[…] as fuel efficiency increases, so do sales,'' after which the system generates a fuel-vs-sales chart. The system thus silently contributes to the exploration by generating charts backing the conversation. Similarly, the art exhibition \citetitle{ephemera} generates images of creatures based on visitors' utterances, capturing their use of taboo language and, hence, externalizing the public's mood.

\textbf{Sense-making with widgets.}
% "inksight knownet vizability vizchat"
%
\citetitle{inksight} facilitates editing natural language explanations of chart findings using GUI widgets. Factual sentences can be created, deleted, moved, and grouped to construct an explanation of chart insights. Data facts in the text are also highlighted and linked with the corresponding chart data points, allowing users to back facts with data by following hyperlinks.

\textbf{Sense-making with spatial interaction.}
% "aras avatar handtracking"
%
\citetitle{inksight} also supports the documentation of chart findings via a combination of sketch-based selection with LLM-generated explanations. After selecting data points on the chart using gestures, the LLM generates a textual insight, which can then be refined by the user. \citetitle{smartboard} also relies on sketching to allow human users to communicate spatial information to an LLM, combined with other modalities such as natural language and charts (\autoref{fig:visreprs2} \textbf{(1)}).

% inksight (sketch)

%Summary for Task 5
\begin{summaryboxLonni}{cyan7}
LLM-augmented visualization systems commonly support sense-making through textual explanations of user queries. The robustness is often enhanced via retrieval-augmented generation or finetuning. Beyond text, several approaches leverage visualization itself as the explanatory medium. They aim at facilitating natural-language questions to be answered through dynamically generated visual encodings (\eg \cite{articulatepro}). Fewer systems combine verbal and visual explanations, although work such as \citetitle{voice} demonstrates the potential of coordinated spoken and visual responses for multimodal sense-making. Recent advances in speech-to-speech foundation models (end-to-end speech model, AudioPaLM \cite{rubenstein_audiopalm_2023}) seem to further suggest that voice-driven interaction paradigms may become increasingly prevalent. Together, these approaches outline a growing and still little-explored design space for explanation in LLM-augmented data visualization systems, spanning textual, visual, and multimodal strategies.
\end{summaryboxLonni}

\begin{comment}
text chatbots provide explanations, which are made more robust with rag or finetuning.

explanations to NL questions can also be provided through visualization, like in \cite{articulatepro}. Only voice combines a spoken and visual explanation \cite{voice} 

expect speech-to-speech models be more widely used soon. AudioPaLM was introduced in june 2023. \cite{rubenstein_audiopalm_2023}.
\end{comment}

% Summary for all task-interaction.
\begin{summaryboxLonni}[LLM-Enabled Tasks]{olive7}

In summary, LLM-enabled interaction has been studied for all tasks in the visualization pipeline. Natural language input facilitates efficient data retrieval, data transformation and visual encoding. The two most common applications are visualization recommendation and exploration of knowledge graphs. The spoken form of natural language input has not been used much yet, despite evidence of its usefulness in specific operational scenarios \cite{leon_talking_2023}, \eg{} sterile environments \cite{aras}, in particular for navigation and sense-making. It is likely that this modality currently does not integrate well with current workflows of experts conducting data analysis \cite{besancon:tel-01684210,isenberg:hal-01095454,wang:hal-02053969,wang:hal-02442690}. As such, focusing solely on spoken interaction with LLMs for visual data analysis could hinder the adoption of new systems and should be considered carefully.

Other interactive inputs, such as spatial inputs and GUI inputs, can also be facilitated by LLMs, for tasks where natural language is less effective. Past research suggests that direct interaction is more effective for fine-grained operations, while natural language facilitates exploration with unspecific outcomes \cite{masson_directgpt_2024}. LLMs can also facilitate \emph{deictic interaction}, such as performing a selection with the mouse or hand gestures followed by a manipulation of that selection with a natural language query \cite{shen_prompting_2025,kim_datahand_2021}. Still, as is apparent in our task-in\-ter\-ac\-tion matrix (\autoref{table:matrix_task_interact}), such uses of LLMs remain anecdotal and could be explored further.

% VisRec, KGs,
% // this goes into the sentence about applications.
% Uses of LLMs for broad tasks in visualization have been explored, applications some applications more extensively (Data science, GIS, Operative surgery) while other aspects remain relatively under-explored (AR/VR, domain-specific applications).

% Stuff about Spoken interaction ill-suited maybe goes here? Something along the lines of: spoken-based interaction with LLMs has been argued to be particularly relevant and interesting for specific use cases in which users cannot use their hands (\eg{} sterile environments \cite{}, manufacturing, ...). However, using speech to interact with a software may also be a nuisance or a concern in shared office spaces \cite{jahncke2012open,Zulfikar} and it is likely that this modality currently does not integrate well with current workflows of experts conducting data analysis \cite{besancon:tel-01684210,isenberg:hal-01095454,wang:hal-02053969,wang:hal-02442690}. As such, focusing solely on spoken-based interaction with LLMs for visual data analysis could hinder the adoption of developed solutions and should be considered carefully. 
\end{summaryboxLonni}

%% --------------------------
\subsection{\textcolor{yellow7!60!black}{Visual representation}}
\label{sec:representations}

LLM-based approaches have been used for a variety of visual representations (2D charts, spatial representations, net\-work\discretionary{/}{}{/}graphs, other representations), as we review next.

%In this part, we review the visual representations used. Below we distinguish 2D charts, spatial representations, network / graphs, and other representation kinds.

\subsubsection{2D charts}

Traditional visualization techniques for tabular data are the most commonly supported visual representations in LLM systems. In this category we group scatterplots, line charts, bar charts (including stacked, histograms etc.), and pie charts.
\NumReprChart{}\,/\,\NumSurveyed{} papers employ visualizations of 2D charts. Their prevalence is due to their widespread use in visual analytics, but also because they are easily produced and understood by LLMs: the underlying data can be stored in structured text formats such as CSV or JSON, and visualizations can be simply generated using popular implementation frameworks (\autoref{sec:data-format}).

\subsubsection{Spatial visualizations}

Spatial visual representations (maps, scalar or vector fields, density plots) are frequently employed and concern \NumReprSpatial{}\,/\,\NumSurveyed{} papers in our corpus. They are more complex to generate or manipulate than charts by LLMs as the data-to-visual mapping is often not straightforward. Focus regions, for instance, can be at a larger scale than individual data points (\eg{}  clusters, and spatial relationships).

2D maps are a common representation, used notably in Geographic Information Systems (GIS). Five papers \cite{autonomousgis,citygpt,lasek,chatgrid,vist5} describe LLM-enabled GIS systems.
In \citetitle{chatgrid}, to answer queries, the LLM generates spatial SQL queries (using PostGIS), which returns entries linked to visual features for the system to display on screen. This way does not treat spatial information, only semantics, while the visualization system displays relevant information.
In the volumetric visualization category we found three papers that tackle molecular visualization \cite{voice,handmol,genius} and three papers that tackle anatomical visualization in VR/AR \cite{avatar,humanatlas,aras}.

\subsubsection{Networks and graphs}

\NumReprNetwork{}\,/\,\NumSurveyed{} papers study graph and network visualizations, of which the most common are knowledge graphs (KGs) \cite{kosa,c5,knownet,graphologue,chatweaver}. We found two main applications among KGs: the construction and visualization of mindmaps \cite{tailormind,c5,sensecape,graphologue} and the visualization of the knowledge retrieval process \cite{kosa,knownet,chatweaver}. Combining LLM information retrieval through KGs with a visualization can facilitate the establishment of an understanding of the source of the data as well as navigating the thought process in a structured way. Another application of graphs is the visualization of the data transformation process \cite{waitgpt,autonomousgis}. Through an interactive solution graph, \citetitle{waitgpt} facilitates the iterative refinement of the generated visualization by tweaking the node attributes directly. Two additional papers employ graph visualization for domain-specific applications, namely a power grid visualization \cite{citygpt} and a protein interaction network \cite{genevic}.

% Although our survey corpus doesn't contain it, there exists a body of literature studying \emph{generation} of networks/graphs using LLMs \cite{}.

\subsubsection{Other visual representations}

% Idk if I should keep this since so few papers target this.
% I suspect that the systematic search does not capture these papers since it's not "visualization" in the traditional sense.
Visualization systems involving images or videos are rarer. In an image, useful information does not appear by reading individual pixels: only with the whole picture the complex human visual system can make sense of data. Advances in computer vision, and recently multimodal vision LLMs, could offer new perspectives: image-based chart reading \cite{detecting_misleading_vis}, labeling and commentary of images and videos, with concrete applications such as real-time vision-based guiding in 3D space \cite{3dllava}.

Several papers \textcite{c5,inksight} employ hypertext techniques and other variations of textual display (\ie{} text visualization). Being the raw format supported by LLMs, such visualizations can efficiently augment LLM-generated output, \eg{} for explainability \cite{xnli}.

\begin{summaryboxLonni}{yellow7}
Traditional charts are studied by a majority of papers (\NumReprChart{}\,/\,\NumSurveyed{}), followed by network and graphs (\NumReprNetwork{}) which were all knowledge graphs; spatial representations (\NumReprSpatial{}) in particular maps and AR applications, and other representations (\NumReprUnknown{}), including text. Out-of-the-box, LLMs perform well on simple tasks involving charts. Other representations seem to require more training and careful prompting of the LLM. Overall, only a very small subset of the papers we surveyed targeted volumetric visualization applications, probably because of the inherent complexity of their manipulation or creation, often combined with the lack of standard software or library to display and analyze them.
\end{summaryboxLonni}

%% --------------------------
\subsection{\textcolor{brown7!60!black}{Application domain and users}}
\label{sec:domain}

We now discuss how specific challenges that may arise map to specific application domains based on our classification (\autoref{sub:application_domain_justification}). Some papers did not target a specific domain (\eg{}  dealing with charts in general)---we categorized these as belonging to the data science category, as we discuss first.
%\todo{I don't have much to say that is specific to an application domain. And/Or the application domain categories are too broad.}
%\textcolor{orange}{Lonni: if that is the case, then that means that our categories are not good or not important... is that really the case? Reading below I thought that I actually learn quite a bit from this.}

\subsubsection{Data science}

% inksight nl4dvllm kosa cellsync waitgpt dataformulator chartgpt citygpt insightlens meqa interchat datadive lida xnli ava visualizationary

A sizeable quantity of papers (\NumTaskEncoding{}) are Natural Language Visualization Recommender Systems (VRS), \ie{} systems that automatically or semi-automatically generate visual representations from natural language prompts and data. These systems are targeted at data scientists and analysts, with the aim of streamlining a repetitive workflow, but can also facilitate exploratory data analysis by providing a quicker feedback loop between the user intent and the visual response. Here interaction comes into play: the verbal interaction complements more traditional approaches. LLMs excel at summarizing and parsing abstract queries, but are less effective in selection tasks or view manipulation, where traditional GUI-based techniques are better suited. VRS were studied before the advent of LLM \cite{survey_towards_vrs}, and is a well-established field connected to sub-topics such as Question-Answering (QA) and Visual Question-Answering (V-QA), Data Mining, and Information Retrieval.

\subsubsection{Scientific subject domains}

% knownet aras voice handmol genevic
% autonomousgis vist5 citygpt

Two of the surveyed papers directly address molecular visualization in 3D space \cite{voice,handmol}, one focuses on genetic data \cite{genevic}, and a last one on operative surgery \cite{aras}. Scientific disciplines are characterized by rich domain-specific constraints on visualization. For instance, in operative surgery, the use of AI facilitates multitasking; in particular speech-enabled systems can free up the hands for synchronous and sterile manipulations. 
Likewise, \textcite{word2wave} describe \citetitleonly{word2wave}, an NLI for mission programming of autonomous underwater vehicles, and justify the need for a NLI by the need to reconfigure the mission on-the-fly. Typical interfaces are too complex to operate in dynamic environments, and NLIs can render user interfaces more user-friendly.

In macro-molecular visualization, structures of interest often exist at multiple scales. A multi-scale navigation system thus must show data at multiple levels of detail while retaining a global context \cite{scaletrotter}. \citetitle{voice} addresses this point by representing the scene as a tree, where the root is the entire macromolecule and children nodes are sub-components. By pruning this hierarchical structure around the viewpoint, the system can effectively craft a LLM prompt for navigation that contains only spatially relevant information, which works around the limited LLM context window.

To provide a final example, \textcite{vist5} tackle climate data exploration and explain that current natural lan\-guage-based visualization systems adapt poorly to domain-specific da\-ta---in this case geographic data---and are better suited for ge\-ne\-ral-pur\-pose data science tasks as explained above. Indeed, LLMs are well trained on tabular data, traditional charts, and general analytics tasks but perform worse on do\-main-spe\-ci\-fic software, data with spatial relationships, or sensitive data. The \citetitle{vist5} system can adapt to multiple do\-main-spe\-ci\-fic visualization libraries by facilitating the injection of domain-specific data through RAG. 

% \textcolor{orange}{The systems developed in these areas provide functions taht are specific to the needs of the domain area that they target, with, for instance, the possibility to manipulate specific visualization widgets such as cutting planes \cite{voice}, or SOMETHING ELSE \cite{}. In these cases, WHAT ARE SOME OF THE CHALLENGES THEY HAD?}

\subsubsection{Education and social sciences}

% vizchat vizability c5 tailormind aicommentator voice ephemera wander2

While data science and scientific applications dominate, several authors explore visualization that targets audiences such as in education \cite{voice,handmol}, improving accessibility \cite{vizability}, entertainment \cite{aicommentator}, or arts \cite{ephemera,wander2}, which all have their own specific need for LLM-based approaches. In \citetitle{vizability}, the authors explore the topic of chart accessibility, exploring alternative ways to read charts for visually impaired users, \eg{} with keyboard input or natural language questions. With their \citetitleonly{aicommentator} system, \textcite{aicommentator} augment football viewing with an LLM-ge\-ne\-ra\-ted live commentary, relying on knowledge retrieval. In educational settings, LLMs provide a user-friend\-ly interface to the visualization without needing training to get started. In artistic settings, LLMs combined with image generation \cite{ephemera,wander2} expand the potential of visitor participation in art exhibits through generation or adaptation of content for different people. Several authors also emphasize that NLIs facilitate user onboarding \cite{leva}, engagement \cite{aicommentator}, or learning visual analytics \cite{vizchat} due to their ease of use for novices.

\begin{summaryboxLonni}{brown7}
Overall, we observed domain-specific constraints and the related challenges and opportunities for LLM-enabled interaction with visualization. Uses of LLMs for broad tasks in visualization has been explored, for some applications more extensively (Data science, GIS, Operative surgery), while other aspects have received relatively little attention (AR/VR, domain-specific applications). Using natural language to interface users and LLM-assisted interaction with the visualization can prove useful in operational scenarios where individuals have limited access to classical interaction devices. In collaborative work scenarios, LLMs facilitate natural interaction between multiple users and an agent through spoken language \cite{articulatepro}. In public scenarios (considering, \eg{} educational contexts), conversational interactions with a visualization are likely to foster engagement and improve accessibility \cite{aicommentator,mapio}.
% \textcolor{orange}{
% Categorizing papers by application domains is helpful to try and see of the challenges that may be inherent to those. System dealing with STEM are, for instance, more likely to require specific 3D manipulation capabilities considering that a lot of the disciplines covered include spatial and/or volumetric datasets.}
\end{summaryboxLonni}

%% --------------------------
\section{\textcolor{azure7!60!black}{LLM Integration in visualization systems}}
\label{sec:llm}

To effectively connect an LLMs with a visualization system, the LLM's model needs to be conditioned for its task (training, fine-tuning, prompt engineering), while the visualization engine needs to pass data to and from the model.

%We now review, from a technological perspective, how LLM systems and visualization systems can be connected. For the LLM, it involves conditioning the model for its task (training, fine-tuning, prompt engineering). For the visualization, it involves passing visualization data to and from the model.

\subsection{Choice of LLM model}

Generally, one can train a bespoke LLM, fine-tune a pretrained model, or prompt a general-purpose model. The model itself can be cloud-based or local. Training a model from scratch, however, is usually not a cost-effective option. The earliest-family LLM models, BERT and GPT (2018--2022), were closed-source models by Google and OpenAI. When GPT-3 and, later, ChatGPT were released to the public, interest in this technology rose in both the public and academic circles. Back then, people had no option other than these models, and it is thus not surprising that the vast majority of papers (\the\numexpr \NumModelGptthree + \NumModelGptthreefive +\NumModelGptfour + \NumModelGptfouro + \NumModelGptfive \relax) use OpenAI's GPT models. Today's LLM ecosystem is more diverse, including both private (Anthropic's Claude, Google's Gemini, Mistral) and open models (Facebook's LLaMa and, more recently, DeepSeek), including cloud-based models and smaller models that can run locally on consumer hardware.

Interestingly, in our paper pool most authors do not state the reason for their model choice or simply say that OpenAI's models are the state-of-the-art \cite{ava,autonomousgis} at time of writing. \citetitle{vist5,chartgpt} use models in the FLAN-T5 family (\citetitle{chartgpt} uses a fine-tuned version of Flan-T5 ``as it has undergone pre-training on various tasks and possesses strong reasoning capabilities''). \citetitle{word2wave} uses an LLM to generate training data for a \emph{T5-small} seq2seq model. There is little comparison of performances between different LLMs, although some authors mention it as future work \cite{cellsync,vizability}.

\subsection{Prompt engineering}

Conversational LLMs are models trained to simulate a conversation between a user and an artificial assistant. These models are ge\-ne\-ral-pur\-pose and can tackle different tasks without prior training. Instructing the model is done through the system prompt, which describes the system and its role, and the user prompt, to start the conversation. There are several strategies for constructing a system prompt \cite{kim_prompt_strategies_2023}. Most commonly, the prompt contains examples of valid user $\leftrightarrow$ assistant interactions. The terms \emph{zero-shot} and \emph{few-shot} refer to the number of examples of conversation turns provided to the model in the system prompt: a model can have zero or a few examples of conversation turns \cite{brown_few_shot_2020}. In ge\-ne\-ral-pur\-pose LLM use, few-shot improves the model alignment with the desired output, without requiring fine-tu\-ning of the model \cite{rajkumar_eval_text2sql_2022}, corroborated in applications of LLMs in visualization \cite{wang_visualization_2025}. \NumPromptFewShot{} papers from our corpus report employing few-shot. \citetitle{vizability} selects the best matches for the few-shot based on co\-sine-si\-mi\-la\-ri\-ty with the query, an approach similar to RAG, and \citetitle{meqa} plans it as future work. But \textcite{chartgpt} decided against using few-shot, as it was too rigid for the ambiguity of the queries, and instead fine-tuned their model with query examples.

An alternative prompting strategy consists of decomposing of the query into smaller tasks, each responsible for addressing one query element. This technique is called Chain-of-Thought (CoT) and significantly improves reasoning capabilities of LLMs on complex tasks by mimicking a step-by-step thought process \cite{wei_cot_2022}. In their paper assessing effectiveness of detecting misleading visualization, \textcite{detecting_misleading_vis} found that CoT prompting was the most efficient way of dealing with their goal. This method is used by 5 papers \cite{interchat,smartboard,cellsync,chartgpt,smartmlvs}.

\subsection{Data formats}
\label{sec:data-format}

% \textcolor{red}{TODO: how is data represented in LLM prompts. Queue to next part: other data formats than text.}

The primary input for LLMs is natural language, but they can also parse other textual forms of information such as code (often Python), structured data (commonly JSON), data tables (mainly CSV), and rich text (\eg{} Markdown). An essential component to prompt engineering is managing the size of the context window, and data formats have various degrees of performance and token footprint. There is insufficient previous research analyzing data formats across different models, but preliminary results indicate that the format has a sizable impact on the performance, varies strongly across models, and models are sensitive to subtle changes in formatting \cite{long_llms_2025,sclar_quantifying_2024}. We encourage future work to explore this topic applied to visualization recommendation tasks. Compressing data to fit the context window can also be achieved by \emph{pruning}, \ie{} contextually removing irrelevant information based on heuristics. \textcite{chartgpt} mention that they include only the header, column type, and top two rows of table data. \textcite{voice} build the prompt based on the location of the viewport in the scene tree and include only information from nearby branches.

\subsection{Multi-modal LLMs}
\label{sec:multimodal}

Today's latest LLM models support varied input and output, not only text: images, audio, and video.
We found that researchers did not consider LLM-based image generation to be particularly good or effective at producing visual representations, except perhaps for artistic purposes \cite{kouts_lsdvis_2023,wood2022beyond}.

Vision Language Models (VLMs) provide richer reasoning capabilities and resilience in LLM-dri\-ven visualization \cite{mena_augmenting_2025}. For example, in \citetitle{vizchat} a VLM receives a screenshot of the visualization for each query. The screenshot helps the LLM to understand to which part of the visualization the user refers and what content is displayed. \textcite{visualizationary} also employ VLMs for visual reading, but combine that with information gathered from computer vision techniques such as eye tracking prediction, optical character recognition, analysis of color usages, etc. 
\textcite{ava} found that VLMs were capable of iterative visual refinement of a produced representation through a feedback loop, albeit with high latency. Their evaluation highlights that current VLMs are best suited for natural images, but struggle with detecting detailed and localized structures. They call for further work in VLMs and visualization applications, as ``these mul\-ti-mo\-dal foundation models hold[] the potential to fundamentally transform the way we think about visualization and user interaction.'' \textcite{autonomousgis} calls for models trained on spatial data, \ie{} Large Spatial Model (LSM). They postulate that, since LLMs have acquired language skills through training on text corpora, similar models could be trained on spatial data to acquire spatial understanding. Furthermore, such training data is readily available in large quantities.

\subsection{Multi-agent systems}

% \textcolor{orange}{Lonni: Mention what it is again here and what it could bring then examples of papers, then what is surprisingly not done for instance.}

A common approach with LLMs consists in creating multiple agents, each specialized for one task. This ``divide and conquer'' approach---known as \emph{Mixture of Agents} (MoA) \cite{wang_mixture--agents_2024}---improves reliability for complex tasks and reduces the context window by including only relevant information for each prompt. 
\citetitle{voice}, for instance, uses a main \emph{manager agent} to interface between a user and the specialized agents (\emph{explorer, pilot, encyclopedia, guardian}). Likewise, \citetitle{vizchat} employs a \emph{context agent} with access to the pre-stored knowledge about the visualization. The agent is responsible for asking for clarifications in case of ambiguity and to craft a prompt incorporating the instruction and RAG-retrieved data into an \emph{explanation agent}. This two-agent system clearly separates between intent understanding using context and retrieval, on one hand, and explanation generation, on the other hand.

\subsection{Long-term memory}
\label{sec:longterm}

Persistent storage can be added to the system to facilitate persistent post-de\-ploy\-ment learning from past interactions and to maintain consistency across sessions. Several authors applied terminology from cognitive psychology to LLM long-term memory \cite{wheeler_procedural_2025}. We thus distinguish \emph{semantic}, \emph{procedural}, and \emph{episodic} memory. Systems using retrieval from knowledge graphs \cite{vizchat} rely on \emph{semantic} memory---they store and retrieve factual information. \textcite{vist5} found that semantic memory allows the LLM to easily be adapted to different domains and software. \citetitle{vist5} uses long-term memory to store \emph{procedural} knowledge: users can store do\-main-spe\-ci\-fic in\-put-ac\-tion pairs to describe how to perform an action given a query. During subsequent queries, relevant stored information is provided to the LLM as few-shot examples in the prompt. Relevance is scored similar to other RAG approaches, with the cosine similarity between the embedding of the user query and the stored input. \citetitle{autonomousgis} uses \emph{episodic} memory, containing past que\-stion-an\-swer pairs to allow the model to recall past interactions. The authors add that they plan to add a procedural memory system, which would need to contain the solution graph but which is not easily stored as textual key-value pairs.

% \textcolor{orange}{Lonni: Mention what it is again here and what it could bring then examples of papers, then what is surprisingly not done for instance.}

\begin{summaryboxLonni}{azure7}
Model conditioning is key to LLM integration in vis systems, including prompt engineering (few-shot, chain-of-thought), fine-tuning, using mixture of agents, and data input and output formatting.
A vast majority of the papers we surveyed made use of OpenAI's family of models, as we show in \autoref{table:matrix_task_interact}. This limits our capacity to transfer findings to other implementations. LLM technology is constantly evolving: new model capabilities are emerging. Vision and speech models are now widespread, while LLMs supporting video and spatial data are actively being developed. The software ecosystem around LLMS is also constantly mutating, such as with the development of many MCP servers connecting LLMs to existing software. These advancements are an academic opportunity, but researchers in the field should be on the lookout for technology advancements and strive to produce generalizable contributions.
\end{summaryboxLonni}

% \newpage
%%%%%%%%%%%%%%%%%%%%%%%%%%%%%
%% EVALUATION
%%%%%%%%%%%%%%%%%%%%%%%%%%%%%
\section{\textcolor{blue7!60!black}{Evaluation of LLM-Vis systems}}
\label{sec:eval}

Evaluation is a critical aspect of visualization research that has long garnered our methodological attention as a field. Considering the variability of LLM-generated content, we believe that conducting proper evaluation when combined with data visualization is essential for the newly developed tool to eventually be used. But the  evaluation of LLMs is a growing field that has yet to reach maturity and, as such, we believe that surveying current practices in our field is likely to inform on interesting approaches or missed opportunities. 
We thus now review how authors evaluated their LLM-aug\-men\-ted systems, discussing two dimensions: \emph{User evaluation} are reports on external participants (not authors) interacting with the system, whereas \emph{system evaluations} include author-based and automated evaluations (\eg{} benchmarks). \NumEvalUser{} papers contain some form of user evaluation, and \NumEvalSys{} contain system evaluations. Only \NumNoEval{} system papers are not evaluated, which we consider to be preliminary work.

Evaluating LLM-enabled systems is generally difficult due to the stochastic nature of LLMs. In contrast to traditional software, the same ``input'' does not always produce the same ``output,'' LLMs are unpredictable \cite{pimentel2024beyond,Liu_model,song-etal-2025-good,thomas2026jagged} even given the same model. And as LLMs are also a novel technology, evaluation methods are still being developed and results are constantly evolving \cite{Chang_eval,pimentel2024beyond}. Automated model benchmarks were shown to be poorly aligned with human evaluation \cite{liu_g-eval_2023}. This problem is accentuated by the closed-source nature of many cloud-based LLMs, which are continuously and sometimes silently updated \cite{angermeir2025reflections,siddiq2025large}.
The unpredictability is exacerbated by the ambiguity of natural language. Many user queries, if abstract, contain ambiguities that can only be resolved with an understanding of the context and cultural background of the asking person. Both issues (the unpredictability of LLM outputs and the ambiguity of natural language) mean that in non-trivial cases there does not exist a 1-to-1 mapping from a query to a action. Consequently, holistic quantitative evaluations are hard to perform if not impossible, and qualitative evaluations with user feedback are needed.

%% --------------------------
\subsection{User evaluation}

% \citeauthor{survey_isenberg_vis_eval_2013} defines 8 \emph{scenarios} of evaluation \cite{survey_isenberg_vis_eval_2013,survey_lam_infovis_eval_2012}: 

\subsubsection{Study participants}

We collected the number of participants and their levels of expertise. Papers reporting on a user study include an average of \AvgParticipants{} in their experiments. Two papers did not communicate the number \cite{humanatlas,aras}. The level of expertise of participants was as follows: \NumUserExpert{}\,/\,\NumUserEval{} include expert users, \NumUserKnowledgeable{}\,/\,\NumUserEval{} include knowledgeable users, \NumUserNovice{}\,/\,\NumUserEval{} include novice users, and \NumUserMixed{}\,/\,\NumUserEval{} were mixed.

In their 2013 survey on user evaluation in visualization, \textcite{survey_isenberg_vis_eval_2013} observed an increase in the amount of papers including a user evaluation, exceeding 50\% in 2012. The median number of participants was 9. \textcite{survey_merino_vis_eval_2018} similarly found that software visualization system papers included an evaluation in with a median of 13 participants. We found similar figures for user evaluation (\NumUserEval{}\,/\,\NumSurveyed{} papers) and number of participants (mean \AvgParticipants{}, median \MedParticipants{})---consistent with the overall trend in visualization. 

\subsubsection{Protocols and instruments}

Almost all authors employ a similar study structure. It consists of three principal phases: observation of participants' use of the system (either unsupervised or with a precise goal), a semi-structured interview with the moderators, and (a) post-study questionnaire(s).

\subsubsection{Metrics and data collection}

Typical user evaluation in visualization includes both objective performance metrics and empirical preference metrics \cite{survey_lam_infovis_eval_2012}. %User performance is measured with metrics such as task completion time and accuracy. User preference metrics are evaluated with instruments such as questionnaires and interviews.
Focusing on performance metrics, authors evaluate their system with objective scores such as measuring tasks completion time (\NumEvalUserTime{}\,/\,\NumUserEval{}), per\-for\-mance\discretionary{/}{}{/}accu\-ra\-cy (\NumEvalUserPerf{}\,/\,\NumUserEval{}), and recall (\NumEvalUserRecall{}\,/\,\NumUserEval{}).

User preference is evaluated with instruments such as questionnaires and interviews. In three papers, authors report perceived workload with the NASA TLX questionnaire (\cite{waitgpt,voice,datadive}). In five papers authors assess perceived usability with the SUS questionnaire \cite{aicommentator,voice,datadive,word2wave,mapio}. Likert-scale questionnaires are used in 19 studies, where the most common questions assessed preference, ease-of-use, effec\-tive\-ness\discretionary{/}{}{/}use\-ful\-ness, helpfulness, satisfaction, enjoyment, trustworthiness, or reliability.
%
% For qualitative evaluation, studies include post-study questionnaires semi-structured interviews with participants, and authors report qualitative feedback in the paper.

%% --------------------------
\subsection{System evaluation}

Apart from user evaluation, three common means of evaluation are the use of benchmarks of standardized tasks, use-case evaluation performed by the authors, and non-structured qualitative feedback in short papers. Authors of \NumEvalSys{} papers in our sample evaluate their system in such a way. In the ML community, frameworks to evaluate LLM systems are still being developed, \eg{} measuring conciseness \cite{ghafari_concise_2025}, retrieval augmented generation capabilities \cite{survey_yu_rag_2025}, bias detection \cite{fanous_syceval_2025,kumar_no_2025}. Frameworks tailored to visualization applications also exist, such as for visualization recommendation \cite{zeng_evaluation-focused_2022} and reading \cite{bench_pandey_literacy_2025}. To our knowledge, there is no framework for evaluating interaction in visualization.

%% --------------------------
\subsubsection{Benchmarks and standard datasets}
% 3dllava, chartgpt, chatgrid, tailormind, nl4dv-llm, meqa, vizability

Authors of \NumEvalSysBench{} papers evaluate their system with a benchmark from visualization, HCI, or ML. Benchmarks are system-agnostic collections of inputs and expected outputs, which can be used both as training data to improve the model quality or as a metric for evaluation. Several benchmarks exist for visualization \mbox{recommendation}. The benchmark by \textcite{bench_nvbench} (and its derivatives \cite{luo_nvbench_2025,chartgpt}) is used in our corpus most frequently: in 2 papers for evaluation \cite{nl4dvllm,smartmlvs} and in 2 papers for training \cite{vist5,chartgpt}. It comprises 25,000+ natural language utterances $\leftrightarrow$ visualization pairs covering traditional chart types (bar, pie, scatter; as Vega-Lite \cite{vega-lite} specifications).

Several authors produced their own data sets to test their system. \citetitle{voice} contains 92 in-scope tasks and 41 out-of-scope tasks curated by the authors. Building a bespoke test dataset for each paper, however, is time-consuming. This approach is not effective at comparing systems with each other; only a standardized benchmark can do that---it would also be more neutral and exhaustive.

Other benchmarks and dataset exist but were not used by any of the surveyed papers, and could be relevant for some of the applications: from the visualization community \cite{bench_quda,bench_plotlyplus,bench_srinivasan_collecting_2021}, from the ML community \cite{tang-etal-2023-vistext}, and specifically for the task of visual ques\-tion-ans\-wer\-ing \cite{bench_kahou_figureqa_2017,bench_kafle_dvqa_2018,bench_methani_plotqa_2019}.

Off-the-shelf LLMs have also been evaluated for their visualization literacy by \textcite{bench_pandey_literacy_2025} and \textcite{bench_bendeck_empirical_2025}. They found that state-of-the-art models in 2025 (GPT-4, Claude, Gemini, Llama) are efficient at basic chart interpretation but perform poorly on misleading visualization. Other authors studied the trustworthiness, fairness, and bias of LLM-generated and LLM-interpreted visualization \cite{koonchanok_trust_your_gut_2025,kaufman_unfair_2025,kim_chatgpt_advice_2025,alexander_misleading_2024}.

Several authors regret the lack of benchmark datasets fitting their needs. \textcite{vizability}, \eg{} call for the development of inclusive benchmarks that incorporate the viewpoints of blind and low-vision individuals. They also regret the lack of fine-grained metrics in benchmarks, including multi-hop reasoning, queries requiring prior context, etc. \textcite{detecting_misleading_vis} add that there is no benchmark for vision-LLM visualization tasks, being urgently needed.

%``our study revealed the need for a more extensive and inclusive benchmarking dataset that incorporates viewpoints of blind and low-vision individuals'' \cite{vizability}
%they regret the lack of more advanced query and chart types, \eg{} requiring multi-hop reasoning; questions with context; vision; other metrics (fluency, informativeness, relevance)

\subsubsection{Qualitative data collection}

Due to the implicit nature of LLMs and visualization tasks, comparing a result against a baseline is not always feasible nor desirable. \textcite{voice} include both in-scope and out-of-scope queries in their testing to compare how the system reacts in predictable and unpredictable scenarios. In in-scope cases, results could be classified as \emph{correct} or \emph{incorrect}, and obtained a score of 90\% \emph{correct}. In out-of-scope cases, LLM interactions were sorted as \emph{partially correct}, \emph{state-dependent}, or \emph{incorrect} and obtained 43\% or \emph{partially correct}. The authors also tried repeating the same queries to get a glimpse of the model variability on ambiguous questions. This approach seems to be effective for evaluating the system in unpredictable contexts. It shows, however, that a significant amount of responses were incorrect or contextually incorrect, hinting at the lack or robustness and trustworthiness of the model. Perhaps the system needs to be trained to self-report its uncertainty in such cases. Then, an evaluation of reported uncertainty vs.\ correctness could be performed.

\begin{summaryboxLonni}[Evaluation of LLM-Vis systems]{blue7}
In the absence of well-established evaluation frameworks and benchmarks tailored to the specific case of LLM-aug\-men\-ted visualization and interaction, most surveyed papers rely on existing practices from the visualization and HCI communities, which both have a long history of evaluation and establishing evaluation standards \cite{bertini_quality_2011,santos_evaluation_2014,elmqvist_patterns_nodate,survey_isenberg_vis_eval_2013,bertini_quality_2011,survey_merino_vis_eval_2018,}. User evaluations are the dominant approach and typically combine quantitative usability measures---most commonly SUS and NASA-TLX---with qualitative feedback collected through post-study interviews. Per\-for\-mance-ori\-en\-ted metrics such as task completion time, accuracy, or retention are used less frequently and are often task-spe\-ci\-fic. While these evaluations provide insight into usability and user experience, they rarely assess aspects that are, nonetheless, central to LLM-based systems such as perceived correctness, trustworthiness, or graded correctness when handling ambiguous or unspecified queries (\eg{} \cite{voice}). Considering the non-de\-ter\-mi\-ni\-stic nature of LLMs and the fact that they hallucinate, it seems that the deployment and use of LLM-augmented visualization systems is likely to depend on how the systems are perceived by users in those terms, which is rarely considered.

System-level evaluations without user involvement are comparatively rare. Several benchmarks exist---such as NVBench \cite{bench_nvbench} and NL2VIS-de\-rived datasets---but only a subset is used in practice and comparisons across systems remain uncommon. Some papers construct bespoke datasets that serve both as training or fine-tuning material and as an evaluation corpus. While this approach facilitates targeted testing, it limits comparability across systems and raises concerns regarding neutrality, coverage, or generalizability. It seems that preliminary efforts toward benchmark-based evaluation frameworks for visualization recommendation (\eg{} \cite{chen_viseval_2024,bench_nvbench}) have been made, but their adoption remains limited.

In our survey we found no benchmark, dataset, or evaluation framework specifically designed to assess the interaction with LLM-aug\-men\-ted visualization systems. Visualization interaction techniques are thus rarely compared systematically, and evaluation strategies remain fragmented. Prior work has highlighted this lack of evaluation methodology in visualization recommendation more broadly \cite{zeng_evaluation-focused_2022}, a gap that is even more pronounced for interactive and multimodal LLM-based systems. Moreover, even within traditional visualization research the developed systems are rarely assessed in the wild or in their intended application field beyond simplified scenarios or a few examplars \cite{besancon_spatial_interfaces_2021}. The challenges of evaluating LLM-aug\-men\-ted system thus encompass those for both LLM and visualization technologies, and perhaps even further ones.

    \begin{comment}
    in absence of well established frameworks and benchmarks tailored for the applications that we survey, current papers follow general practices in the vis+hci communities, i.e. a user eval to test usability of the system, often with SUS/TLX or similar for quant data, followed by qualitative feedback in interviews. Metrics more specific to the topic that would be interesting to evaluate: perceived trust/correctness of LLM output, degrees of output correctness when handling ambiguous queries \cite{voice}

    system evaluations (non-user studies) are rarer. We identified many potential benchmarks and datasets that were not used in the surveyed papers. Several papers resort to making their own dataset, which doubles down as an evaluation + training/finetuning dataset. Preliminary work regarding both benchmarks and eval framework exist, for visrec \cite{chen_viseval_2024,bench_nvbench}.

    There is no benchmark/dataset/framework for interaction.

    visrec systems are seldomly compared against each other, which makes comparison of different approaches difficult. \cite{zeng_evaluation-focused_2022}

    \end{comment}
\end{summaryboxLonni}

% \newpage
%%%%%%%%%%%%%%%%%%%%%%%%%%%%%
%% OPPORTUNITIES
%%%%%%%%%%%%%%%%%%%%%%%%%%%%%
\section{Research opportunities and discussions}
\label{sec:opportunities}

Our survey of the convergence of LLMs and visualization for interactive data analysis highlights several important research opportunities and challenges, as well as interesting discussion points to consider for our field, which we highlight below.

\subsection{Opportunities for future research}

We first highlight the interesting avenues for future research that stem from our literature survey. 

\subsubsection{The potential of multimodality}

Consumer-grade multimodal models are now capable of reading and generating images and audio. Speech-to-speech models, combined with a voice activity detection mechanism, facilitate low-latency speech conversations with LLM assistants. Previous LLM-based voice assistants included speech-to-text and text-to-speech intermediate conversion steps, which introduced latency and lost precious information conveyed through the voice tone. Image reading and generation capabilities will improve the visual literacy of LLMs \cite{bench_pandey_literacy_2025} by allowing them to take screenshots of the visualization, hence allowing the model to validate the correctness of its own actions and take corrective measures, as shown in \citetitle{3dllava}. Combination of visual and sensory inputs suggest that it will be possible to increase virtual agents' surrounding awareness with ambient microphones and cameras in the room, which should increase the realism of anthropomorphic agents, which can acknowledge and guide users of an interactive visualization.

While images and audio are most common, other LLM models develop different modalities. 3D LLMs \cite{hong_3d-llm_2023,3dllava} can treat 3D point-cloud scenes as input and have improved spatial reasoning capabilities, beyond Vision Language Models \cite{hong_3d-llm_2023}. Video LLMs are able to treat image sequences as output and reason about temporal information \cite{liu_st-llm_2025}. Combined with intention prediction systems, LLMs can become \emph{agentic} and \emph{proactive}, \ie{} initiate actions without explicit user prompts. Hence, current research is developing au\-dio\discretionary{/}{}{/}vi\-deo streaming models, which bring LLMs closer to human interaction patterns \cite{wang_videollm_2025,chen_videollm-online_2024,wang_simul-whisper_2024}.   

\subsubsection{Interfaces with external data sources}

Agentic LLM systems have the capability to interact with external systems through \emph{tool calls}. The Model Context Protocol (MCP; \href{https://modelcontextprotocol.io}{\texttt{modelcontextprotocol\discretionary{}{.}{.}io}}) format offers a standardized API for LLMs to interact with external tools. Adoption of this ecosystem will facilitate an easy and rapid integration of LLM agents into existing visualization software. Furthermore, it will allow LLMs to act as a \emph{middleware}, connecting multiple unrelated systems with little effort. Proprietary LLM products such as Google Gemini already rely on tool calls to give language models access to the file system or searching the web, which facilitates data retrieval from unknown sources. Yet, we have not seen such capabilities used in visualization research. Opening access to web-sourced data in visualization will expand the possibilities of exploratory data analysis, where the data analyzed may be complemented and compared with online databases. Automated retrieval of online content poses challenges for automatic assessment and communication about data reliability. Such uncertainty can be communicated visually, orally, or in written form, and modalities for communicating uncertainty influence people's perception, \eg{} spoken uncertainty leads to riskier decisions, while text leads to lower confidence \cite{stokes_voicing_2024}.

\begin{comment}
    see sentence above: and address it.
    
     For example, the survey \cite{survey_deldjoo_genrecsys_2024} coined the term Gen-RecSys for retrieval systems with generative capabilities.

\end{comment}

\subsubsection{Tailoring visualization and interaction to users}

LLM-based interaction offers new capabilities for capturing user intent and resolving ambiguity, creating unprecedented opportunities to make data visualization more accessible through intuitive interfaces aligned with users’ needs. This is important since personalized experiences can increase user engagement, knowledge transfer, and reduce cognitive load, which are key to an effective scientific \mbox{understanding} and communication \cite{koc2022connecting,schonborn2016nano,Ynnerman:exp}. Further research is needed to understand the mechanisms underlying these experiences with LLM-powered visualization, as well as to develop more personalized interactions tailored to individual users. Several factors may contribute to this goal. To initiate this discussion, we suggest the following directions.

\textbf{Personal interaction histories.}
With the development of long-term memory mechanisms (\autoref{sec:longterm}), LLMs gained the ability to recall information from previous sessions. Combined with user identification \cite{wang_diarizationlm_2024}, LLMs learn from and about users and provide more relevant, engaging, and accessible information in return.\footnote{This venue is already being explored by commercial actors such as OpenAI as indicated in their privacy policies, see, \eg{} \url{https://openai.com/policies/privacy-policy/}.} 

\textbf{Holistic user modelling.}
Interpersonal communication occurs not only through explicit conversation, but also with non-verbal signals such as gestures, facial expressions, tone of voice, posture, and proxemics in collaborative setups \cite{Beyan,mccall2016mapping,McDuff,Williamson}. While some of these signals are already leveraged in collaborative data analysis scenarios and systems \cite{sereno:hal-02971697}, the use of LLMs to interpret these signals is an interesting research direction that has not been extensively explored.

\textbf{Aesthetic qualities.} %  and affect in communication
Aesthetics is, in general, an important aspect of visual design~\cite{hook_somaesthetic_2016}. We would be interested to learn how image generation models and style transfer techniques could be adopted to leverage aesthetics in visualization \cite{kouts_lsdvis_2023,lida}, for example to foster engagement or interest, or to elicit or avoid specific emotions (\eg{} \cite{besancon:hal-01795744,besancon:hal-02381513}) depending on the target audience and communicative goal \cite{wood2022beyond}.

\textbf{Adaptive use of modalities.}
Several authors have characterized input modalities in relation to visualization goals \cite{perrin_vis_interact}. Individual differences need to be taking into account, however, which can stem from background, preference, expertise, and abilities \cite{liu_survey_2020}.  \textcite{vizability} studied adaptations for blind and low vision people, and \textcite{interchat} explored combinations of mouse and natural language interaction, but further work is needed.

In summary, LLM-augmented visualization facilitates the tailoring of interaction and presentation to individual users by capturing intent, resolving ambiguity, and adapting over time, across modalities, capabilities, and contexts. By leveraging personal interaction histories, richer user models, aesthetic qualities, and adaptive use of the available modalities, future systems could reduce cognitive load, improve accessibility, foster better engagement, and better support diverse user goals. To realize these benefits, however, a deeper empirical understanding is needed of how personalization mechanisms influence engagement, comprehension, and trust in visualization and LLM-augmented systems. In parallel, the visualization research community must continue to advance its evaluation strategies to better guide and validate research outcomes.

\begin{comment}
    * long term memory => learning from user patterns and coherent conversations accross sessions
    * reading user proxemics, behavior and non-verbal communication => through extra sensors, which opens the door to ``implicit interaction''
    * contextualize visualization with user's personal experiences => for the affective dimension. Aestetics could play a role too \cite{lida, TODO gen ai style transfer something}, and sensitive content \cite{wood2022beyond} etc.
    * offering multiple alternative interaction modalities make visualization more accessible: there is not "one best interaction for the task", it depends on user preference, expertise, abilities.
\end{comment}

\subsubsection{The challenge of evaluation}

Visualization, as a field, has a positive history of rethinking evaluation methods to ensure the field does not solely focus on quantitative and performance oriented metrics and that our results have a wider impact \cite{survey_lam_infovis_eval_2012}. We often re-evaluate, \eg{} through the BELIV workshop \cite{BELIV_2020} the methods on which we rely for different measures of importance and are open to explore and adopt new methods for measuring or reporting results \cite{besancon:hal-01980268,dragicevic:hal-01377894} and to improve transparency \cite{haroz:hal-01947432,preregBeliv} or reusability of results \cite{isenberg:hal-04697010}. This emphasis stems from the recognition that assessing the impact, adoption, and benefits of visualization is a multifactorial problem, shaped by intertwined cognitive, social, organizational, and contextual factors that inherently resist reduction to a single metric or experimental paradigm. Combining visualization research to LLM research does not simplify the space but introduces new challenges of evaluation \cite{Crisan} that, moving forward, we need to address as a field in order to produce reliable results and systems that can be, eventually, adopted for real-life situation and thus for our field to continue having impact beyond our publications.

% \textcolor{orange}{Lonni: still drafty points from me below, would need to be made into a fully mature version.}
Only a subset of papers provide user evaluation (\NumEvalUser{}\,/\,\NumSurveyed{}) and fewer still have qualified and tried to improve the trustworthiness of LLM-produced information. This is particularly important as LLMs are prone to hallucinations \cite{survey_ji_hallucinations_2023}. It is further important because adoption of LLM-augmented visualization will eventually rely on the trust that users will place in the system. As the adoption of new visualization software is already an open challenge of the visualization research community \cite{wang:hal-02053969}, we see this lack of trust evaluation as a new threat to the wider adoption of the solutions we are likely to develop in the future. We thus argue that a comprehensive evaluation framework of LLM-augmented visualizations needs to be developed that consider all of the aspects that are necessary for a wide adoption of such systems and encompasses, at least, all elements that we have considered in this report. 

Furthermore, authors clearly express the need for benchmarking datasets and we find statements such as ``a benchmark dataset is urgently needed. Our evaluation efforts were limited by two main factors: (1) the absence of a benchmark dataset and (2) the need for refined evaluation metrics'' \cite{detecting_misleading_vis} and ``our study revealed the need for a more extensive and inclusive benchmarking dataset that incorporates viewpoints of blind and low-vision individuals'' \cite{vizability}. Providing the community with useful benchmarking datasets would lead to more systematic, reproducible, and comparable evaluations across systems and allow us to track progress in the field.

% \newpage
%%%%%%%%%%%%%%%%%%%%%%%%%%%%%
%% DISCUSSION
%%%%%%%%%%%%%%%%%%%%%%%%%%%%%

\subsection{Discussion}
\label{sec:discussion}
While conducting this work, we have encountered several challenges that we believe are important to discuss for the visualization community but also for the wider research community.

%\subsubsection{Inclusion of preprints}

As noted in \autoref{sub:preprint_inclusion}, it is commonly advised that survey papers should focus on published and, as such, peer-reviewed contributions only. Doing so, however, could pose significant problems when surveying LLM research considering the extremely fast-moving pace of the field and the fact that many landmark papers in LLM/AI do not even get submitted for peer-review (including technical papers from companies \cite{llama2,gpt4}, the ``Survey of Large Language Models'' which is updated continuously \cite{zhao_survey_2025}, etc.). We consequently decided to include preprints from arXiv in this STAR. We further decided to clearly distinguish those papers from peer-re\-viewed ones through a specific citation marker throughout our paper to indicate that, perhaps, these papers should be considered with more caution. Given the high turnover rate of LLM-related publications, where models, datasets and benchmarks are superseded within months, excluding preprints may risk producing an outdated and scientifically unrepresentative picture of the field by the time the STAR is available. Including them, however, may pose a threat to the reliability of the represented picture of the state of the art. We believe that, in this respect, our approach ensures relevance while still preserving methodological transparency. The approach we adopted is likely the most complete one in terms of inclusion of papers, while maintaining a transparent presentation of the surveyed space, but we want to argue that the PRISMA guidelines \cite{prisma} may have to be updated to consider this general problem, possibly our solution, and perhaps a wider call for other solutions.

In addition, a challenge we encountered is to try and verify whether some of the preprints we identified eventually ended up being published to maintain the most accurate image of the field. We saw such updates happening for several papers that we had initially included as preprints, and we mention these now as published. While arXiv provides a way for authors to link to the peer-re\-viewed version of their preprints, many authors 
%us included (e.g., at the time of writing \cite{voice}), 
do not use this possibility. With LLM-driven research accelerating, systematic mechanisms for tracing preprint development, peer-review status, and potential retractions (see below) will become increasingly pivotal to ensure scientific integrity in future literature surveys such as STARs.

Our survey finally highlighted that it, at times, was difficult to extract the information needed to categorize the papers. In particular, some of the papers we surveyed did not mention the necessary details about their LLM implementation, with some of them not even mentioning which version of a specific LLM they used or whether it was trained or fine-tuned and how. Considering how unreliable LLMs can be \cite{Dale_2021,liu2023summary}, not providing this kind of information further complicates matters. 
We believe that we should, as a field, ensure that we communicate such information in the manuscript to ensure that our research remains both transparent and replicable. This is even more important considering that LLM-oriented research is already difficult to replicate \cite{bench_bendeck_empirical_2025}.
The frequent lack of implementation details is especially problematic in the context of LLM-enabled visualization systems, where subtle differences (\eg{} prompt templates, versioning) can lead to varying outputs. Without precise reporting, findings may become non-replicable, and comparisons across different systems unreliable, thus hindering the cumulative progress of the field. We thus argue for the adoption of clear reporting standards for future LLM-Visualization research, which will not only improve reproducibility \cite{Cockburn_2020,isenberg:hal-04697010} but also facilitate meaningful meta-analysis in forthcoming STARs.

% \newpage
%%%%%%%%%%%%%%%%%%%%%%%%%%%%%
%% CONCLUSION
%%%%%%%%%%%%%%%%%%%%%%%%%%%%%
\section{Conclusion}
We have surveyed the existing work in employing LLMs in data visualization systems for enabling interaction with the data. First, following the PRISMA methodology, we classified the literature in the field according to four axes: LLM-enabled visualization tasks, interaction modalities, visual representations, and application domain. Second, we contribute a survey of implementation and evaluation of LLM-enabled visualization systems to help establish research standards for future work. Third, we discussed opportunities for future work in the field at the intersection between LLMs, data visualization, and interaction. In closing, LLM-enabled interaction is changing the landscape of visualization. While it is expanding access as part of driving a paradigm shift, it is also exposing critical limitations in reliability and evaluation. Our systematic STAR synthesis clarifies current capabilities, gaps, and methodological challenges and offers a foundation for the future advancement of robust multimodal visio-verbal systems that actively support rich, transparent, and trustworthy human-computer interaction.
% \newpage
%%%%%%%%%%%%%%%%%%%%%%%%%%%%%
%% ACKNOWLEDGEMENTS
%%%%%%%%%%%%%%%%%%%%%%%%%%%%%
\section{Acknowledgments}

This work was partially supported by Marcus and Amalia Wallenberg Foundation (2023.0128) and (2023.0130), and Knut and Alice Wallenberg Foundation (2019.0024).
We are grateful to the authors who provided the figures used as illustrations in this report. We also wish to thank I. Viola for his feedback on the manuscript and fruitful discussions during the data collection process.

\section{Author contributions}

We list below the contributions of each individual author based on the CRediT author statement (\url{https://credit.niso.org/}).%\vspace{-1em}

\credit{M. Brossier}{%
    1, % Conceptualization
    1, % Data curation
    1, % Formal analysis
    0, % Funding acquisition
    0, % Investigation
    1, % Methodology
    0, % Project administration
    0, % Resources
    0, % Software
    0, % Supervision
    0, % Validation
    1, % Visualization
    1, % Writing -- original draft
    1  % Writing -- review & editing
}%

\credit{T. Isenberg}{%
    1, % Conceptualization
    1, % Data curation
    1, % Formal analysis
    0, % Funding acquisition
    0, % Investigation
    1, % Methodology
    0, % Project administration
    0, % Resources
    0, % Software
    1, % Supervision
    0, % Validation
    0, % Visualization
    1, % Writing -- original draft
    1  % Writing -- review & editing
}%

\credit{K. Schönborn}{%
    1, % Conceptualization
    0, % Data curation
    0, % Formal analysis
    1, % Funding acquisition
    0, % Investigation
    0, % Methodology
    1, % Project administration
    0, % Resources
    0, % Software
    1, % Supervision
    0, % Validation
    0, % Visualization
    1, % Writing -- original draft
    1  % Writing -- review & editing
}%

\credit{J. Unger}{%
    1, % Conceptualization
    1, % Data curation
    1, % Formal analysis
    1, % Funding acquisition
    0, % Investigation
    0, % Methodology
    0, % Project administration
    0, % Resources
    0, % Software
    1, % Supervision
    0, % Validation
    0, % Visualization
    0, % Writing -- original draft
    1  % Writing -- review & editing
}%

\credit{M. Romero}{%
    0, % Conceptualization
    1, % Data curation
    0, % Formal analysis
    0, % Funding acquisition
    0, % Investigation
    0, % Methodology
    0, % Project administration
    0, % Resources
    0, % Software
    0, % Supervision
    0, % Validation
    0, % Visualization
    0, % Writing -- original draft
    1  % Writing -- review & editing
}%

\credit{J. Björklund}{%
    1, % Conceptualization
    1, % Data curation
    0, % Formal analysis
    0, % Funding acquisition
    0, % Investigation
    1, % Methodology
    0, % Project administration
    0, % Resources
    0, % Software
    0, % Supervision
    0, % Validation
    0, % Visualization
    0, % Writing -- original draft
    1  % Writing -- review & editing
}%

% \credit{I. Viola} {
%     0, % Conceptualization
%     0, % Data curation
%     0, % Formal analysis
%     0, % Funding acquisition
%     0, % Investigation
%     0, % Methodology
%     0, % Project administration
%     0, % Resources
%     0, % Software
%     0, % Supervision
%     0, % Validation
%     0, % Visualization
%     0, % Writing -- original draft
%     0  % Writing -- review & editing
% }

\credit{A. Ynnerman}{%
    1, % Conceptualization
    0, % Data curation
    0, % Formal analysis
    1, % Funding acquisition
    0, % Investigation
    1, % Methodology
    1, % Project administration
    0, % Resources
    0, % Software
    1, % Supervision
    0, % Validation
    0, % Visualization
    1, % Writing -- original draft
    1  % Writing -- review & editing
}%

\credit{L. Besançon}{%
    1, % Conceptualization
    1, % Data curation
    1, % Formal analysis
    1, % Funding acquisition
    0, % Investigation
    1, % Methodology
    1, % Project administration
    0, % Resources
    0, % Software
    1, % Supervision
    0, % Validation
    0, % Visualization
    1, % Writing -- original draft
    1  % Writing -- review & editing
}%

{\footnotesize \insertcreditsgranular}%

%\newpage
\printbibliography
\balance

\end{document}

%% file: statistics.tex
\newcommand{\NumSurveyed}{48}

\newcommand{\NumReprChart}{24}

\newcommand{\NumReprSpatial}{18}
\newcommand{\NumReprNetwork}{18}
\newcommand{\NumReprUnknown}{0}

\newcommand{\NumModelGptthree}{1}
\newcommand{\NumModelGptthreefive}{12}
\newcommand{\NumModelGptfour}{11}
\newcommand{\NumModelGptfouro}{4}
\newcommand{\NumModelGptfive}{0}

\newcommand{\NumPromptFewShot}{9}

\newcommand{\NumUserEval}{25}
\newcommand{\AvgParticipants}{13.8}
\newcommand{\MedParticipants}{12}
\newcommand{\NumUserExpert}{8}
\newcommand{\NumUserKnowledgeable}{12}
\newcommand{\NumUserNovice}{3}
\newcommand{\NumUserMixed}{5}

\newcommand{\NumNoEval}{4}
\newcommand{\NumEvalUser}{25}
\newcommand{\NumEvalSys}{28}

\newcommand{\NumEvalSysBench}{7}

\newcommand{\NumEvalUserTime}{7}

\newcommand{\NumEvalUserPerf}{10}
\newcommand{\NumEvalUserRecall}{1}

\newcommand{\NumIdentifiedPapers}{353}

\newcommand{\NumIdentifiedPapersOpportunistic}{8}
\newcommand{\NumIdentifiedPreprintsArxiv}{78}

\newcommand{\NumFilteredPapers}{240}
\newcommand{\NumCodedPapers}{48}
\newcommand{\NumCodedPreprints}{3}

\newcommand{\NumTaskEncoding}{22}

%% file: coding_summary.tex
\SetTblrInner{colsep=3pt,rowsep=0pt}

% \begin{tblrtikzbelow}
% \path[pattern color=gray9,pattern=checkerboard,
% draw=blue3, ultra thick, rounded corners]
% (table.north west) rectangle (table.south east);
% \end{tblrtikzbelow}%

\begin{tblr}{
    rows = {ht=\baselineskip}, % make rows fixed height
    hlines = {dotted},
    hline{1,2,3,47} = {solid},
    vlines = {dotted},
    vline{1,2,6,10,15,19,24,31} = {solid},
}

\SetCells{c}

% header row 1
papers (\NumCodedPapers{}) &
\SetCell[c=4]{c,brown7!50} {\hypersetup{hidelinks}\hyperref[sec:domain]{appl. domain}} & & & &
\SetCell[c=4]{c,yellow7!50} vis. repr. & & & &
\SetCell[c=5]{c,olive7!50} vis. task & & & & & 
\SetCell[c=4]{c,cyan7!50} interaction & & & &
\SetCell[c=5]{c,azure7!50} {\hypersetup{hidelinks}\hyperref[sec:llm]{LLM}} & & & & &
\SetCell[c=7]{c,blue7!50} {\hypersetup{hidelinks}\hyperref[sec:eval]{eval.}} & & & & & & \\

% header row 2 rotated
\SetRow{cmd=\rotatebox{90}}
& % this is the leftmost empty cell on row 2
data sci. \& math. & med., bio. \& chem. & phys. \& eng. & edu. \& social sci. & 
charts & custom & spatial & network &
data retrieval & data transf. & visual encoding & sense-making & navigation &
NL text & NL speech & spatial interaction & mouse widgets &
OpenAI & Llama & Gemini & other & LLM-vision &
num. participants & particip. expertise & system eval & has benchmark & has case study & has comp. study & has empir. study \\

%% Content generated from the spreadsheet. Copy-paste below!
%% https://docs.google.com/spreadsheets/d/1ST1xs4iCY97ktxx0VDT-uvtoHrX9Adr6nbC3DrPFKKc/edit?gid=1356180918#gid=1356180918

InkSight \hfill \cite{inksight} & \SetCell{brown7} & & & & \SetCell{yellow7} & \SetCell{yellow7} & & & & & & \SetCell{olive7} & & & & \SetCell{cyan7} & \SetCell{cyan7} & \SetCell{azure7!50}3.5 & & & & & \SetCell{blue7!50}12 & \SetCell{blue7!50}K & & & & &  \\   
Autonomous GIS \hfill \cite{autonomousgis} & & & \SetCell{brown7} & & \SetCell{yellow7} & & \SetCell{yellow7} & \SetCell{yellow7} & & \SetCell{olive7} & \SetCell{olive7} & & & \SetCell{cyan7} & & & & \SetCell{azure7!50}4 & & & & &  &  & \SetCell{blue7} & & \SetCell{blue7} & &  \\   
HINTs \hfill \cite{hints} & \SetCell{brown7} & & & & & & \SetCell{yellow7} & & & & & & & & & & &  & & & & &  &  & & & & &  \\   
NL4DV-LLM \hfill \cite{nl4dvllm} & \SetCell{brown7} & & & & \SetCell{yellow7} & & & & & \SetCell{olive7} & \SetCell{olive7} & & & \SetCell{cyan7} & & & & \SetCell{azure7!50}4 & & & & &  &  & \SetCell{blue7} & \SetCell{blue7} & & \SetCell{blue7} &  \\   
VizChat \hfill \cite{vizchat} & & & & \SetCell{brown7} & \SetCell{yellow7} & & & & & & & \SetCell{olive7} & \SetCell{olive7} & \SetCell{cyan7} & & & \SetCell{cyan7} & \SetCell{azure7!50}4 & & & & \SetCell{azure7} &  &  & \SetCell{blue7} & & & & \SetCell{blue7} \\   
KOSA \hfill \cite{kosa} & \SetCell{brown7} & & & & \SetCell{yellow7} & & & \SetCell{yellow7} & & & & & & \SetCell{cyan7} & & & &  & & & \SetCell{azure7} & &  &  & \SetCell{blue7} & & & \SetCell{blue7} & \SetCell{blue7} \\   
CellSync \hfill \cite{cellsync} & \SetCell{brown7} & & & & & \SetCell{yellow7} & & & & & & \SetCell{olive7} & & \SetCell{cyan7} & & & & \SetCell{azure7!50}3 & & & & & \SetCell{blue7!50}20 & \SetCell{blue7!50}E & & & & &  \\   
WaitGPT \hfill \cite{waitgpt} & \SetCell{brown7} & & & & & & & \SetCell{yellow7} & & \SetCell{olive7} & \SetCell{olive7} & & & \SetCell{cyan7} & & & \SetCell{cyan7} & \SetCell{azure7!50}4 & & & & & \SetCell{blue7!50}12 & \SetCell{blue7!50}K & \SetCell{blue7} & & & \SetCell{blue7} &  \\   
VizAbility \hfill \cite{vizability} & & & & \SetCell{brown7} & \SetCell{yellow7} & & & & & & & \SetCell{olive7} & \SetCell{olive7} & \SetCell{cyan7} & & & \SetCell{cyan7} & \SetCell{azure7!50}4 & & & & \SetCell{azure7} & \SetCell{blue7!50}6 & \SetCell{blue7!50}K & \SetCell{blue7} & \SetCell{blue7} & & \SetCell{blue7} &  \\   
C5 \hfill \cite{c5} & & & & \SetCell{brown7} & & & & \SetCell{yellow7} & & & & \SetCell{olive7} & & & & & \SetCell{cyan7} & \SetCell{azure7!50}3.5 & & & & & \SetCell{blue7!50}8 & \SetCell{blue7!50}K & & & & &  \\   
KnowNet \hfill \cite{knownet} & & \SetCell{brown7} & & & & & & \SetCell{yellow7} & \SetCell{olive7} & & & \SetCell{olive7} & & \SetCell{cyan7} & & & \SetCell{cyan7} & \SetCell{azure7!50}4 & & & & &  &  & \SetCell{blue7} & & \SetCell{blue7} & &  \\   
Wander 2.0 \hfill \cite{wander2} & & & & \SetCell{brown7} & & & \SetCell{yellow7} & & & & & \SetCell{olive7} & & \SetCell{cyan7} & & & &  & & & & &  &  & & & & &  \\   
Data Formulator \hfill \cite{dataformulator} & \SetCell{brown7} & & & & \SetCell{yellow7} & & & & & \SetCell{olive7} & \SetCell{olive7} & & & \SetCell{cyan7} & & & \SetCell{cyan7} & \SetCell{azure7!50}3.5 & & & & & \SetCell{blue7!50}10 & \SetCell{blue7!50}E & & & & &  \\   
VIST5 \hfill \cite{vist5} & & & \SetCell{brown7} & & \SetCell{yellow7} & & \SetCell{yellow7} & & & \SetCell{olive7} & \SetCell{olive7} & & \SetCell{olive7} & \SetCell{cyan7} & & & \SetCell{cyan7} &  & & & \SetCell{azure7} & & \SetCell{blue7!50}24 & \SetCell{blue7!50}K & & & & &  \\   
ChartGPT \hfill \cite{chartgpt} & \SetCell{brown7} & & & & \SetCell{yellow7} & & & & & \SetCell{olive7} & \SetCell{olive7} & & & \SetCell{cyan7} & & & \SetCell{cyan7} & \SetCell{azure7!50}3 & & & \SetCell{azure7} & & \SetCell{blue7!50}12 & \SetCell{blue7!50}K & \SetCell{blue7} & \SetCell{blue7} & & &  \\   
CityGPT \hfill \cite{citygpt} & & & \SetCell{brown7} & & \SetCell{yellow7} & & \SetCell{yellow7} & & & \SetCell{olive7} & \SetCell{olive7} & \SetCell{olive7} & & \SetCell{cyan7} & & & &  & & & & &  &  & \SetCell{blue7} & & \SetCell{blue7} & \SetCell{blue7} &  \\   
ARAS \hfill \cite{aras} & & \SetCell{brown7} & & & & & \SetCell{yellow7} & & & & & \SetCell{olive7} & & & \SetCell{cyan7} & \SetCell{cyan7} & & \SetCell{azure7!50}3.5 & & & & & \SetCell{blue7!50}? & \SetCell{blue7!50}E & & & & &  \\   
Tailor-Mind \hfill \cite{tailormind} & & & & \SetCell{brown7} & & & & \SetCell{yellow7} & \SetCell{olive7} & & & \SetCell{olive7} & & \SetCell{cyan7} & & & & \SetCell{azure7!50}4 & & & \SetCell{azure7} & & \SetCell{blue7!50}24 & \SetCell{blue7!50}K & \SetCell{blue7} & \SetCell{blue7} & & &  \\   
InsightLens \hfill \cite{insightlens} & \SetCell{brown7} & & & & \SetCell{yellow7} & & & & & & & \SetCell{olive7} & \SetCell{olive7} & \SetCell{cyan7} & & & & \SetCell{azure7!50}4 & & & & & \SetCell{blue7!50}12 & \SetCell{blue7!50}E & \SetCell{blue7} & & & \SetCell{blue7} &  \\   
ChatWeaver \hfill \cite{chatweaver} & \SetCell{brown7} & & & & & & & \SetCell{yellow7} & \SetCell{olive7} & & & \SetCell{olive7} & & & & & &  & & & & & \SetCell{blue7!50}10 & \SetCell{blue7!50}M & & & & &  \\   
MapIO \hfill \cite{mapio} & & & & \SetCell{brown7} & & & \SetCell{yellow7} & & & & & & & & & & &  & & & & & \SetCell{blue7!50}10 & \SetCell{blue7!50}M & & & & &  \\   
 \hfill \cite{humanatlas} & & \SetCell{brown7} & & \SetCell{brown7} & & & \SetCell{yellow7} & & & & \SetCell{olive7} & \SetCell{olive7} & \SetCell{olive7} & & & \SetCell{cyan7} & \SetCell{cyan7} &  & & & & & \SetCell{blue7!50}? & \SetCell{blue7!50}M & & & & &  \\   
3D-LLaVA \hfill \cite{3dllava} & \SetCell{brown7} & & & & & & \SetCell{yellow7} & & & & \SetCell{olive7} & \SetCell{olive7} & & \SetCell{cyan7} & & & &  & & & \SetCell{azure7} & \SetCell{azure7} &  &  & \SetCell{blue7} & \SetCell{blue7} & & \SetCell{blue7} & \SetCell{blue7} \\   
LASEK \hfill \cite{lasek} & & & \SetCell{brown7} & & & & \SetCell{yellow7} & & & & \SetCell{olive7} & & & \SetCell{cyan7} & & & &  & & \SetCell{azure7} & & &  &  & \SetCell{blue7} & & \SetCell{blue7} & &  \\   
MEQA \hfill \cite{meqa} & \SetCell{brown7} & & & & \SetCell{yellow7} & & & & \SetCell{olive7} & \SetCell{olive7} & \SetCell{olive7} & & & \SetCell{cyan7} & & & \SetCell{cyan7} &  & & & & &  &  & \SetCell{blue7} & \SetCell{blue7} & & \SetCell{blue7} &  \\   
Smartboard \hfill \cite{smartboard} & & & & \SetCell{brown7} & & \SetCell{yellow7} & & & & \SetCell{olive7} & \SetCell{olive7} & \SetCell{olive7} & & \SetCell{cyan7} & & \SetCell{cyan7} & \SetCell{cyan7} & \SetCell{azure7!50}4 & & & & \SetCell{azure7} & \SetCell{blue7!50}8 & \SetCell{blue7!50}E & \SetCell{blue7} & & \SetCell{blue7} & &  \\   
InterChat \hfill \cite{interchat} & \SetCell{brown7} & & & & \SetCell{yellow7} & & & & & \SetCell{olive7} & \SetCell{olive7} & & \SetCell{olive7} & \SetCell{cyan7} & \SetCell{cyan7} & \SetCell{cyan7} & \SetCell{cyan7} &  & & & \SetCell{azure7} & \SetCell{azure7} & \SetCell{blue7!50}10 & \SetCell{blue7!50}K & \SetCell{blue7} & & \SetCell{blue7} & &  \\   
Ephemera \hfill \cite{ephemera} & & & & \SetCell{brown7} & & \SetCell{yellow7} & & & & & & \SetCell{olive7} & & & \SetCell{cyan7} & & &  & & & & &  &  & & & & &  \\   
Sensecape \hfill \cite{sensecape} & \SetCell{brown7} & & & & & & & \SetCell{yellow7} & & \SetCell{olive7} & & & & \SetCell{cyan7} & & & & \SetCell{azure7!50}4 & & & & &  &  & & & & &  \\   
Graphologue \hfill \cite{graphologue} & \SetCell{brown7} & & & & & & & \SetCell{yellow7} & & & & \SetCell{olive7} & \SetCell{olive7} & \SetCell{cyan7} & & & & \SetCell{azure7!50}4 & & & & & \SetCell{blue7!50}7 & \SetCell{blue7!50}E & \SetCell{blue7} & & & &  \\   
AiCommentator \hfill \cite{aicommentator} & & & & \SetCell{brown7} & \SetCell{yellow7} & \SetCell{yellow7} & & & & & & \SetCell{olive7} & & \SetCell{cyan7} & & & & \SetCell{azure7!50}3.5 & & & & & \SetCell{blue7!50}16 & \SetCell{blue7!50}M & \SetCell{blue7} & & & &  \\   
DataDive \hfill \cite{datadive} & \SetCell{brown7} & & & & \SetCell{yellow7} & & & & & & & \SetCell{olive7} & \SetCell{olive7} & \SetCell{cyan7} & & & & \SetCell{azure7!50}4 & & & \SetCell{azure7} & & \SetCell{blue7!50}21 & \SetCell{blue7!50}N & \SetCell{blue7} & & & & \SetCell{blue7} \\   
Avatar \hfill \cite{avatar} & & \SetCell{brown7} & & & & & \SetCell{yellow7} & & & & & \SetCell{olive7} & & & \SetCell{cyan7} & \SetCell{cyan7} & &  & & & & &  &  & & & & &  \\   
ChatGrid \hfill \cite{chatgrid} & \SetCell{brown7} & & \SetCell{brown7} & & & & \SetCell{yellow7} & \SetCell{yellow7} & \SetCell{olive7} & & & & \SetCell{olive7} & \SetCell{cyan7} & & & &  & & & & &  &  & \SetCell{blue7} & \SetCell{blue7} & & &  \\   
GENIUS \hfill \cite{genius} & & & \SetCell{brown7} & & & & \SetCell{yellow7} & & & \SetCell{olive7} & \SetCell{olive7} & & & & \SetCell{cyan7} & \SetCell{cyan7} & & \SetCell{azure7!50}4-o & \SetCell{azure7} & & & &  &  & & & & &  \\   
SmartMLVs \hfill \cite{smartmlvs} & \SetCell{brown7} & & & & \SetCell{yellow7} & & & & & \SetCell{olive7} & \SetCell{olive7} & & \SetCell{olive7} & \SetCell{cyan7} & & & & \SetCell{azure7!50}4 & & & & & \SetCell{blue7!50}32 & \SetCell{blue7!50}K & \SetCell{blue7} & & & &  \\   
Word2Wave \hfill \cite{word2wave} & & & \SetCell{brown7} & & & \SetCell{yellow7} & & & & & \SetCell{olive7} & & & & \SetCell{cyan7} & & &  & & & \SetCell{azure7} & & \SetCell{blue7!50}15 & \SetCell{blue7!50}N & \SetCell{blue7} & & & \SetCell{blue7} &  \\   
VOICE \hfill \cite{voice} & & \SetCell{brown7} & & \SetCell{brown7} & & & \SetCell{yellow7} & & & & & \SetCell{olive7} & \SetCell{olive7} & \SetCell{cyan7} & \SetCell{cyan7} & & & \SetCell{azure7!50}4 & & & & & \SetCell{blue7!50}12 & \SetCell{blue7!50}N & \SetCell{blue7} & & & &  \\   
LIDA \hfill \cite{lida} & \SetCell{brown7} & & & & \SetCell{yellow7} & & & & & \SetCell{olive7} & \SetCell{olive7} & & & \SetCell{cyan7} & & & & \SetCell{azure7!50}3.5 & & & & \SetCell{azure7} &  &  & \SetCell{blue7} & & & &  \\   
XNLI \hfill \cite{xnli} & \SetCell{brown7} & & & & \SetCell{yellow7} & & & & & & & & & \SetCell{cyan7} & & & &  & & & & & \SetCell{blue7!50}12 & \SetCell{blue7!50}M & \SetCell{blue7} & & & \SetCell{blue7} &  \\   
LEVA \hfill \cite{leva} & \SetCell{brown7} & & & & \SetCell{yellow7} & & \SetCell{yellow7} & & & & & & & & & & &  & & & & & \SetCell{blue7!50}20 & \SetCell{blue7!50}K & \SetCell{blue7} & & \SetCell{blue7} & &  \\   
HandMol \hfill \cite{handmol} & & \SetCell{brown7} & & & & & \SetCell{yellow7} & & & & & & & & & & &  & & & & &  &  & & & & &  \\   
GENEVIC \hfill \cite{genevic} & & \SetCell{brown7} & & & \SetCell{yellow7} & & & \SetCell{yellow7} & \SetCell{olive7} & \SetCell{olive7} & \SetCell{olive7} & & & \SetCell{cyan7} & & & \SetCell{cyan7} & \SetCell{azure7!50}4 & & & & &  &  & \SetCell{blue7} & & \SetCell{blue7} & &  \\   
AVA \hfill \cite{ava} & \SetCell{brown7} & & & & & & \SetCell{yellow7} & & & & \SetCell{olive7} & & & \SetCell{cyan7} & & & \SetCell{cyan7} & \SetCell{azure7!50}4-o & & & & \SetCell{azure7} & \SetCell{blue7!50}4 & \SetCell{blue7!50}E & \SetCell{blue7} & & & \SetCell{blue7} & \SetCell{blue7} \\     

\end{tblr}

%% file: task_interaction_matrix.tex
\newcommand{\ColScale}{
  \begin{tikzpicture}[baseline=-0.5ex]
    \shade[left color=white,right color=green] (0,0) rectangle (2,0.3);
    \draw (0,0.3) node[above]{0\%};
    \draw (2,0.3) node[above]{100\%};
  \end{tikzpicture}
}

\begin{tblr}{
    width={\linewidth},
    rowspec={Q|Q|Q|Q|Q|Q},
    colspec={l|X|X|X|X|X|l},
    hline{1,7} = {2-6}{solid},
    vline{1,8} = {2-5}{solid},
}
 & data retrieval & data transformation & visual encoding & navigation & sense-making \\

%% Content generated from the spreadsheet. Copy-paste below!
%% https://docs.google.com/spreadsheets/d/1ST1xs4iCY97ktxx0VDT-uvtoHrX9Adr6nbC3DrPFKKc/edit?gid=1356180918#gid=1356180918

text &  \SetCell{green!15!white!85} \small\cite{chatgrid} \cite{tailormind} \cite{genevic} \cite{knownet} \cite{meqa} \cite{mirror} \cite{havior} & \SetCell{green!33!white!67} \small\cite{autonomousgis} \cite{chartgpt} \cite{citygpt} \cite{dataformulator} \cite{nl4dvllm} \cite{genevic} \cite{interchat} \cite{lida} \cite{meqa} \cite{mirror} \cite{havior} \cite{sensecape} \cite{smartboard} \cite{smartmlvs} \cite{vist5} \cite{waitgpt} & \SetCell{green!38!white!62} \small\cite{3dllava} \cite{autonomousgis} \cite{ava} \cite{chartgpt} \cite{citygpt} \cite{dataformulator} \cite{nl4dvllm} \cite{genevic} \cite{interchat} \cite{lasek} \cite{lida} \cite{meqa} \cite{mirror} \cite{havior} \cite{smartboard} \cite{smartmlvs} \cite{vist5} \cite{waitgpt} & \SetCell{green!31!white!69} \small\cite{3dllava} \cite{aicommentator} \cite{citygpt} \cite{datadive} \cite{wander2} \cite{tailormind} \cite{graphologue} \cite{insightlens} \cite{knownet} \cite{cellsync} \cite{smartboard} \cite{visualizationary} \cite{vizability} \cite{vizchat} \cite{voice} & \SetCell{green!23!white!77} \small\cite{chatgrid} \cite{datadive} \cite{graphologue} \cite{insightlens} \cite{interchat} \cite{havior} \cite{smartmlvs} \cite{vist5} \cite{vizability} \cite{vizchat} \cite{voice} & \SetCell{green!71!white!29} 34 \\
speech &  \SetCell{green!0!white!100} \small & \SetCell{green!6!white!94} \small\cite{articulatepro} \cite{genius} \cite{interchat} & \SetCell{green!8!white!92} \small\cite{articulatepro} \cite{genius} \cite{interchat} \cite{word2wave} & \SetCell{green!10!white!90} \small\cite{articulatepro} \cite{aras} \cite{avatar} \cite{ephemera} \cite{voice} & \SetCell{green!4!white!96} \small\cite{interchat} \cite{voice} & \SetCell{green!17!white!83} 8 \\
spatial & \SetCell{green!0!white!100} \small & \SetCell{green!6!white!94} \small\cite{genius} \cite{interchat} \cite{smartboard} & \SetCell{green!8!white!92} \small\cite{genius} \cite{interchat} \cite{smartboard} \cite{humanatlas} & \SetCell{green!10!white!90} \small\cite{aras} \cite{avatar} \cite{inksight} \cite{smartboard} \cite{humanatlas} & \SetCell{green!4!white!96} \small\cite{interchat} \cite{humanatlas} & \SetCell{green!6!white!94} 3 \\
widgets & \SetCell{green!6!white!94} \small\cite{genevic} \cite{knownet} \cite{meqa} & \SetCell{green!17!white!83} \small\cite{chartgpt} \cite{dataformulator} \cite{genevic} \cite{interchat} \cite{meqa} \cite{smartboard} \cite{vist5} \cite{waitgpt} & \SetCell{green!21!white!79} \small\cite{ava} \cite{chartgpt} \cite{dataformulator} \cite{genevic} \cite{interchat} \cite{meqa} \cite{smartboard} \cite{vist5} \cite{waitgpt} \cite{humanatlas} & \SetCell{green!15!white!85} \small\cite{c5} \cite{inksight} \cite{knownet} \cite{smartboard} \cite{vizability} \cite{vizchat} \cite{humanatlas} & \SetCell{green!10!white!90} \small\cite{interchat} \cite{vist5} \cite{vizability} \cite{vizchat} \cite{humanatlas} & \SetCell{green!31!white!69} 15 \\
& \SetCell{green!17!white!83} 8 & \SetCell{green!38!white!62} 18 & \SetCell{green!46!white!54} 22 & \SetCell{green!48!white!52} 23 & \SetCell{green!25!white!75} 12 & \\

\end{tblr}